\documentclass[showpacs,preprintnumbers,amsmath,amssymb]{revtex4}

\usepackage{epsfig}% Include figure files
\usepackage{graphicx}% Include figure files
\usepackage{dcolumn}% Align table columns on decimal point
\usepackage{bm}% bold math

\newcommand{\EQ}{\begin{equation}}
\newcommand{\EN}{\end{equation}}
\newcommand{\ea}{\end{eqnarray}}
\newcommand{\ba}{\begin{eqnarray}}

\newcommand{\bear}{\begin{eqnarray}}
\newcommand{\ear}{\end{eqnarray}}

\begin{document}

%%%%%%%%%%%%%%%%%%%%%%%%%%%%%%%%%%%%%%%%%%%%%%%%%%%%%%%%%%%%%%%%%%%%%%%%%%
%                               Title                                    %
%%%%%%%%%%%%%%%%%%%%%%%%%%%%%%%%%%%%%%%%%%%%%%%%%%%%%%%%%%%%%%%%%%%%%%%%%%

\title{Scattering theories for the 1D Hubbard model}
\author{J. M. P. Carmelo}
\affiliation{GCEP - C. Physics, University of Minho, Campus Gualtar, P-4710-057 Braga,
Portugal} \affiliation{Department of Physics, Massachusetts Institute of Technology,
Cambridge, Massachusetts 02139-4307}
\author{K. E. Hibberd}
\affiliation{GCEP - C. Physics, University of Minho, Campus Gualtar, P-4710-057 Braga,
Portugal} \affiliation{Centre for Mathematical Physics, The University of Queensland,
QLD, 4072, Australia}
\author{N. Andrei}
\affiliation{Department of Physics and Astronomy, Rutgers University, Piscataway, NJ
08855}
%\date{21 March 2006}
%\date{\today}
%%%%%%%%%%%%%%%%%%%%%%%%%%%%%%%%%%%%%%%%%%%%%%%%%%%%%%%%%%%%%%%%%%%%%%%%%%
%                              abstract                                  
%%%%%%%%%%%%%%%%%%%%%%%%%%%%%%%%%%%%%%%%%%%%%%%%%%%%%%%%%%%%%%%%%%%%%%%%%%

\begin{abstract}
In one-dimensional (1D) non-perturbative many-electron problems such as the 1D Hubbard
model the electronic charge and spin degrees of freedom separate into exotic quantum
objects. However, there are two different representations for such objects and associated
scattering quantities whose relation is not well understood. Here we solve the problem 
by finding important information about the relation between the corresponding alternative 
choices for one-particle scattering states. Our study reveals why one of these representations, 
the {\it pseudofermion} representation, is the most suitable for the description of the unusual 
finite-energy spectral and dynamical properties of the model. This is a problem of physical 
importance, since the exotic independent charge and spin finite-energy spectral features 
observed by angle-resolved photoelectron spectroscopy in quasi-1D metals was found 
recently to correspond to the charge and spin quantum objects of the pseudofermion 
representation.
\end{abstract}

\maketitle
%%%%%%%%%%%%%%%%%%%%%%%%%%%%%%%%%%%%%%%%%%%%%%%%%%%%%%%%%%%%%%%%%%%%%%%%%%
%                              body of paper                             
%%%%%%%%%%%%%%%%%%%%%%%%%%%%%%%%%%%%%%%%%%%%%%%%%%%%%%%%%%%%%%%%%%%%%%%%%%
\section{INTRODUCTION}
Recent photo-emission experiments in quasi-one-dimensional (1D) 
organic metals \cite{spectral0,spectral,Eric} have revealed finite-energy 
charge and spin spectral features similar to those predicted by the
1D Hubbard model, one of the few realistic electronic models for 
which all the energy eigenstates and their
energies can be exactly calculated \cite{Lieb,Takahashi}. In addition
to describing quasi-1D organic metals the model can be experimentally
realized with unprecedented precision in ultra-cold fermionic atomic
systems and one may expect very detailed experimental results over a
wide range of parameters to be available \cite{Zoller,Chen}.

In contrast to the model finite-energy physics, in the last fifteen years 
the low-energy behavior of correlation functions has been the subject 
of many studies 
\cite{Woy,Ogata,Kawakami,Frahm,Brech,Karlo,93-94}.
Recently, the theoretical description of the experimentally observed 
finite-energy spectral features could be given using the {\it pseudofermion
dynamical theory} (PDT), based on pseudofermionic accounting of
the excitation spectrum \cite{V-1,LE}. For example, the one-electron 
independent charge and spin spectral features experimentally
observed for energies above the broken-symmetry states \cite{BS}
in the metallic phase of low-dimensional organic compounds  
\cite{spectral0} correspond to power-law singularities along
lines whose shape coincides with the energy dispersions of 
the charge and spin pseudofermion scatterers, respectively 
\cite{spectral,S-P0,S-P}. Furthermore, the momentum
and energy dependence of the corresponding
spectral-weight distributions is fully controlled by the
pseudofermion scattering \cite{V-1,LE}.  Also other exotic properties 
of the model  were experimentally observed in low-dimensional complex
materials \cite{properties,super}.

Another accounting of the elementary-excitation scattering quantities
was given in Refs. \cite{Natan,S0,S}. While the latter description,
based on {\it spinon-holon} accounting, was used in many theoretical studies
of electronic correlated systems, its relation to the finite-energy
spectral and dynamical properties is not well understood. The main
goal of this paper is the clarification of the relation between the
two scattering theories, of Refs. \cite{S-P0,S-P} and
Refs. \cite{Natan,S0,S}, respectively, for the 1D Hubbard model, in
view with the application of the former approach to dynamical
calculations.

We begin by briefly reviewing the two approaches. The spin $1/2$ {\em
spinon} and the {\em holon} representation used in
Refs. \cite{Natan,S0,S} was first introduced for the chiral invariant
Gross-Neveu Hamiltonian \cite{Natan0}, a continuum version of the
Hubbard Hamiltonian.  The spin $1/2$ spinon corresponds to the
color spin $1/2$ particle of charge zero, whereas the massless
excitation of that reference describes a quantum object associated
with the charge degrees of freedom, now termed holon. The same
excitation structure, decoupled spinon and holon was found
for the Kondo model \cite{NA80}. Subsequently, the same spinon -
spin $1/2$ spin wave representation was discussed \cite{Faddeev} for
the isotropic Heisenberg model. The reason for the same spin
excitation structures in these models is that in each case their
dynamics is based on a spin exchange interaction, $\vec{S} \cdot
\vec{S'}$, leading to spin Bethe-ansatz (BA) equations that
are isomorphic, differing only in their energy functions (and in their charge
structure). Hence, the spin excitations possess the same quantum
numbers and the same scattering matrices \cite{Natan}. The charge
excitations, on the other hand, differ in the various models and in
particular in the Hubbard model the holons acquire a $\eta=1/2$
quantum number, associated with a charge $SU(2)$ group \cite{S0,S,HL}.

While the spinons and holons of the conventional spinon-holon
representation of Refs. \cite{Natan,S0,S} were
introduced by direct association with specific occupancy
configurations of the quantum numbers of the BA
solution, the pseudofermion description, used below, corresponds to
occupancy configurations of ``rotated electrons'', related to the
original electron by a unitary transformation, chosen so that the
states are characterized by their double occupancy. The
pseudofermion description also refers to well defined occupancy
configurations of the quantum numbers of the BA solution and the
$\eta$-spin and spin $SU(2)$ symmetries \cite{I,IIIb}. Its choice of the
objects whose occupancy configurations describe the energy eigenstates
profits from the transformation laws under the electron -
rotated-electron unitary transformation.  Indeed, within the
pseudofermion theory a well defined set of $\eta$-spin $1/2$ holons
and spin $1/2$ spinons called Yang holons and HL spinons (Heilman-Lieb
spinons), respectively, are invariant under that transformation. As a
result of that invariance such objects are neither scatterers nor
scattering centers \cite{S-P0,S-P}.  All pseudofermion branches except
one are closely related to the $\eta$-spin $1/2$ holons or spin $1/2$
spinons introduced in Ref. \cite{I} in terms of rotated-electron
occupancies, which are different from those of the conventional
spinon-holon representation of Refs. \cite{Natan,S0,S,TS}, 
as further discussed in future sections of this paper. Indeed, the 
remaining holonic and spinonic degrees of freedom beyond the 
above Yang holons and HL spinons give rise to $\eta$-spin-zero 
$2\nu$-holon composite charge $c\nu$ pseudofermions and 
spin-zero $s\nu$ $2\nu$-spinon composite pseudofermions, respectively, 
where $\nu =1,2,3...$.  Finally, the energy eigenstates also involve 
occupancy configurations of charge $c$ or $c0$ pseudofermions, 
which are independent of the holonic and spinonic degrees of 
freedom and thus are $\eta$-spin-less and spin-less objects.

The pseudofermion scattering description is a good starting point for
the derivation of dynamical properties by means of the PDT
\cite{spectral,V-1,LE}. There are three main reasons why that description
is more adapted to the calculations of finite-energy spectral 
functions than the related but different representation of Refs. \cite{I,II} 
and the conventional spinon-holon representation of Refs. \cite{Natan,S0,S}.
First, the pseudofermion energy spectrum has no residual-interaction
terms \cite{IIIb}, in contrast to that of the related objects of Refs. \cite{I,II}. 
This is behind the factorization of the one- and two-electron spectral
functions in terms of pseudofermion spectral functions \cite{V-1}.
Second, the pseudofermion $S$ matrix is a simple phase
factor, whereas that of the holons and spinons of the 
conventional representation of Refs. \cite{Natan,S0,S}
is a matrix of dimension larger than one. Third, for the metallic
phase all one-pseudofermion scattering states of the pseudofermion 
theory correspond to many-pseudofermion energy 
eigenstates, whereas some of the alternative spinon-holon
scattering states of the theory of Refs. \cite{Natan,S0,S}
do not refer to energy eigenstates. As further discussed 
in Sec. IV-B, the second and
third points considerably simplify the calculation of the 
finite-energy spectral functions, when expressed in terms 
of Lehmann representations \cite{V-1}.

The pseudofermion theory is a generalization for finite values of
the on-site repulsion $U$ of the $U/t>>1$ method introduced in
Ref. \cite{Penc}, where $t$ is the first-neighbor transfer
integral. The natural excitation basis that arises for $U/t>>1$ is the
one considered in the studies of Refs. \cite{I,S-P} for all values of
$U/t$.  For instance, the spin-less fermions of
Refs. \cite{Ogata,Penc}, the spins or spinons of
Refs. \cite{Ogata,Penc,PWA}, and the doublons and holons of
Ref. \cite{Tohyama} correspond to the $c0$ pseudofermions, spinons,
and $\eta$-spin projection $-1/2$ and $+1/2$ holons, respectively, of
Refs. \cite{I,S-P} for $U/t>>1$.  The studies of Ref. \cite{LE}
confirm that in the limit of low energy the finite-energy PDT
reproduces the well known results and behaviors of the spectral and
correlation functions previously obtained
\cite{Woy,Ogata,Kawakami,Frahm,Brech,Karlo,93-94} by use of
conformal-field theory \cite{CFT} and bosonization \cite{Bozo}.

Here we show that both the representations are faithful and correspond
to two different choices of one-particle scattering states and thus that there is
no inconsistency between the two corresponding definitions of
scatterers and scattering centers. Moreover, we complete the
preliminary analysis of Ref. \cite{S-P0} and confirm that the
pseudofermion representation is the most suitable for the description
of the finite-energy spectral and dynamical properties. Our results apply to
other integrable interacting problems besides the 1D Hubbard model
\cite{NA80,tj} and therefore have wide applicability.

Concerning the conventional spinon-holon representation, in this paper 
we use the notation of Refs. \cite{S0,S}. We note that
in spite of using a different notation, the spinon-holon
representation of Ref. \cite{Natan} is the same as that used in these
references. The paper is organized as follows: In section II we
introduce the 1D Hubbard model and summarize the basic information
about the pseudofermion description needed for
our investigations. In Sec. III we consider the pseudofermion
scattering quantities and clarify the connection of the
pseudofermion and pseudofermion-hole phase shifts and
corresponding $S$ matrices to the elementary-excitation phase 
shifts and $S$ matrices, respectively, previously obtained in
Refs. \cite{Natan,S0,S,Korepin79}. In Sec. IV we extend the 
spinon-holon conventional scattering
theory of Refs. \cite{Natan,S0,S} to the larger Hilbert subspace of
the pseudofermion scattering theory and confirm the faithful character
of both quantum-object representations. Moreover, in that section we
discuss the suitability for applications to the study of the
finite-energy spectral and dynamical properties. Finally, Sec. V
contains the concluding remarks.

%%%%%%%%%%%%%%%%%%%%%%%%%%%%%%%%%%%%%%%%%%%%%%%%%%%%%%%%%%%%%%%%%%%%%%%%%%
\section{THE 1D HUBBARD MODEL AND THE PSEUDOFERMION SCATTERING THEORY}

In this section we introduce the 1D Hubbard model and summarize the
concepts and results concerning the rotated electrons \cite{I} and the
pseudofermion description \cite{V-1,LE,IIIb} that are needed for our 
studies.

The basic Hamiltonian, defined on a 1D lattice with $N_a$ sites, 
is given by,
\begin{equation}
{\hat{H}}_{H} = -t \sum_{j=1}^{N_a} (c^{\dagger}_{j,\sigma} c_{j+1,\sigma} +h.c.) +U\sum_{j=1}^{N_a}
n_{j\uparrow}n_{j\downarrow} \equiv  \hat{T}+U\,\hat{D} \, ,
\end{equation}
where, the operator
$c_{j,\,\sigma}^{\dagger}$ (and $c_{j,\,\sigma}$) 
creates (and annihilates) a spin-projection $\sigma
$ electron at lattice site $j=1,2,...,N_a$, $\hat{n}_{j,\,\sigma} =
c_{j,\,\sigma }^{\dagger }\,c_{j,\,\sigma }$ counts the number of
spin-projection $\sigma$ electrons at lattice site $j$,
$\hat{T}=-t\sum_{\sigma=\uparrow ,\,\downarrow
}\sum_{j=1}^{N_a}\Bigl[c_{j,\,\sigma}^{\dag}\,c_{j+1,\,\sigma} + h.
c.\Bigr]$ is the {\it kinetic-energy} operator, and $\hat{D} =
\sum_{j=1}^{N_a}\hat{n}_{j,\,\uparrow}\,\hat{n}_{j,\,\downarrow}$ is
the electron double-occupation operator.

The model has an obvious $U(1) \times SU(2)$ symmetry,
\begin{eqnarray}
c_{j, \sigma} & \longrightarrow & e^{i\theta}c_{j, \sigma} \nonumber \\
c_{j, \sigma} & \longrightarrow & U_{\sigma,\sigma' } c_{j,
\sigma'} \, , \nonumber
\end{eqnarray}
expressing the charge conservation and invariance under spin
rotation. The associated generators are given by 
the number operator, \ba {\hat
N}=\sum_{j=1}^{N_a}(\hat{n}_{j\uparrow}+\hat{n}_{j\downarrow}) \, ,  \ea
and the spin operators,
\ba S^z_s=\frac{1}{2}\sum_{j=1}^{N_a}
(\hat{n}_{j\uparrow}-\hat{n}_{j\downarrow}),~~~~S_s^{+}=\sum_{j=1}^{N_a}
c^{\dagger}_{j,\downarrow}c_{j,\uparrow},~~~~ S_s^{-} =(S^+)^{*} \, , \ea 
respectively. There is
another (less obvious) charge SU(2) invariance present in a slightly
modified version of the model \cite{HL}, \ba {\hat{H}}_H-\frac{U}{2}\sum_{j=1}^{N_a}
\Bigl(\hat{n}_{j\downarrow}+\hat{n}_{j\uparrow}-{1\over 2}\Bigr)\, ,\ea where a chemical potential
$U/2$ term was added to the Hamiltonian. In a grand canonical ensemble
the model will be half filled. Equivalently, the symmetry will show up
if we work in the canonical ensemble and choose the filling
appropriately. The symmetry is realized by number density and pair
creation and annihilation operators,
\ba
S^z_c=\frac{1}{2}\sum_{j=1}^{N_a}
(n_{i\uparrow}+n_{i\downarrow}-1),~~~~S^+_c=\sum_{j=1}^{N_a}
(-1)^j c^{\dagger}_{j,\downarrow} c^{\dagger}_{j,\uparrow},~~~~S^{-}_c=(S_c^+)^{*}. 
\ea
As the number operator does not commute with $S_c^{\pm}$, the symmetry
manifests itself only upon comparing excitations in systems with
different number of electrons. For historical reasons we refer to the charge 
$SU(2)$ as $\eta$-spin \cite{HL}.
Adding a chemical potential $\mu$ and magnetic field $H$ the
Hamiltonian takes the form,
\begin{equation}
\hat{H}={\hat{H}}_{SO(4)} + \sum_{\alpha =c,\,s}\mu_{\alpha}\, {\hat{S}}_{\alpha}^z \, ;
\hspace{0.5cm} {\hat{H}}_{SO(4)} = {\hat{H}}_{H} - {U\over 2}\Bigl[\,\hat{N} - {N_a\over 2}\Bigr] \, ;
\hspace{0.5cm}{\hat{H}}_{H} = \hat{T}+U\,\hat{D} \, , \label{H}
\end{equation}
where $\mu_c=2\mu$, $\mu_s=2\mu_0 H$, and $\mu_0$ is the
Bohr magneton. The
momentum operator is given by $\hat{P} =
\sum_{\sigma=\uparrow,\,\downarrow }\sum_{k}\, \hat{n}_{\sigma} (k)\,
k$, where the spin-projection $\sigma$ momentum distribution operator
reads $\hat{n}_{\sigma} (k) = c_{k,\,\sigma }^{\dagger}\,c_{k,\,\sigma }$.

Throughout this paper we use units of both Planck constant $\hbar$ and
lattice constant $a$ one. We denote the electronic charge by $-e$, the
lattice length by $L=N_a\,a=N_a$, and the $\eta$-spin value $\eta$
(and spin value $S$) and $\eta$-spin projection $\eta_z$ 
(and spin projection $S_z$) of the energy eigenstates by $S_c$ and
$S_c^z$ (and $S_s$ and $S_s^z$), respectively. For the description of
the transport of charge in terms of electrons (and electronic holes),
the Hamiltonians provided in Eq.  (\ref{H}) describe $N$ electrons
(and $[2N_a -N]$ electronic holes) in a lattice of $N_a$ sites
\cite{I}.  The Hamiltonian $\hat{H}_{SO(4)}$ given in that equation
commutes with the six generators of the $\eta$-spin and spin $SU(2)$
algebras and has $SO(4)$ symmetry \cite{HL}. In this paper we study
scattering processes that result from ground-state - excited-state
transitions. For simplicity, we consider that the initial ground state
has electronic density $n=N/L$ and spin density
$m=[N_{\uparrow}-N_{\downarrow}]/L$ in the ranges $0\leq n \leq 1$ and
$0\leq m \leq n$, respectively.

Lieb and Wu \cite{Lieb} were able to diagonalize the Hamiltonian by
means of the BA approach and write down equations that allow
the determination of all energy eigenvalues and eigenstates. For a
detailed account see, for example, Ref \cite{Natan} and references
therein, where the accounting of the quantum numbers charaterizing the
states is given in the conventional spinon-holon language. We shall 
choose here a different accounting based on the rotated-electron
holons and spinons mentioned in the previous section. It was shown in Ref. \cite{I}
that all energy eigenstates of the model can be described in terms of
occupancy configurations of rotated electrons for the whole Hilbert
space and for all values of $U/t$. Further, based on rotated-electrons
pseudofermions can be introduced \cite{I,IIIb} which provides a
convenient starting point for a dynamical theory \cite{V-1,LE}.

We now proceed to summarize the pseudofermion description of 
the excitation spectrum. It is closely related to the holons and spinons
as defined in Ref. \cite{I}. Such holons (and spinons) have $\eta$-spin 
$1/2$, with $\eta$-spin projection $\pm 1/2$, and spin zero (and spin $1/2$, 
with spin projection $\pm 1/2$, and no charge degrees of freedom), 
and are defined so that the rotated-electron double occupation content 
equals the number of $-1/2$ holons. Here we use the designations 
$\pm 1/2$ holons and $\pm 1/2$ spinons in terms of the $\eta$-spin and 
spin projections, respectively. Starting from a given ground state, the
pseudofermion subspace (PS) is spanned by the excited energy
eigenstates that can be described in terms of occupancy configurations
of pseudofermions, Yang holons, and HL spinons \cite{I,IIIb}.
The one- and two-electron excitations are contained in the PS
\cite{V-1,IIIb}.

The $c0$ pseudofermions have no spin and
$\eta$-spin degrees of freedom.  The $c\nu$ pseudofermions 
for $\nu > 0$ (and $s\nu$ pseudofermions), are composite objects
having  $\eta$-spin zero (and spin zero) consisting of
an equal number $\nu=1,2,...$ of $-1/2$ holons and $+1/2$ holons
(and $-1/2$ spinons and $+1/2$ spinons).  In this paper we use the
notation $\alpha\nu$ pseudofermion, where $\alpha=c,\,s$ and $\nu
=0,1,2,...$ for the $c\nu$ branches and $\nu =1,2,...$ for the $s\nu$
branches. As further discussed in Sec. IV-A, the holons, $c0$ 
pseudofermions, and composite $c\nu$ pseudofermions are charged 
objects. The different pseudofermion branches correspond to well
known types of BA excitations. For instance, in the PS the $c0$
pseudofermion occupancy configurations describe the BA charge
distribution of $k's$ excitations and those of the $c\nu$
pseudofermions for $\nu > 0$ (and $s\nu$ pseudofermions) 
describe the BA charge string excitations of length $\nu$ 
(and BA spin string excitations of length $\nu$).

The properties of the Yang holons and HL spinons follow from the 
invariance of the three generators of the $\eta$-spin $SU(2)$
algebra and three generators of the spin $SU(2)$ algebra,
respectively, under the electron - rotated-electron unitary transformation.
Indeed, the Yang holons and HL spinons are also
invariant under that transformation \cite{I}. Therefore, the operators 
that transform such objects have the same form in terms of electronic 
and rotated-electron creation and annihilation operators. For instance, 
the $\eta$-spin off diagonal generator that creates (and annihilates) 
an on-site electronic Cooper pair transforms a $+1/2$ Yang holon 
(and a $-1/2$ Yang holon) into a $-1/2$ Yang holon (and
a $+1/2$ Yang holon). Furthermore, the spin off diagonal generator 
that flips an on-site electronic up spin (and down spin) onto 
an on-site electronic down spin (and up spin) also transforms a 
$+1/2$ HL spinon (and a $-1/2$ HL spinon) into a $-1/2$ HL spinon 
(and a $+1/2$ HL spinon). Thus, the occupancies of these objects 
involving Yang holons with different $\eta$-spin projections $+1/2$ 
and $-1/2$ and/or HL spinons with different spin projections 
$+1/2$ and $-1/2$ describe the energy eigenstates that are not 
contained the BA solution. The corresponding energy eigenstates
contained in that solution have precisely the same pseudofermion
occupancy configurations and the same Yang holon
and HL spinon total numbers. However, all the Yang holons
and HL spinons of the latter states have the same $\eta$-spin
and spin projections, respectively. (For more information about Yang 
holons and HL spinons see Sec. 2.4 of Ref. \cite{I}.)

We denote the number of $\alpha\nu$ pseudofermions 
by $N_{\alpha\nu}$ and the number
$\pm 1/2$ Yang holons ($\alpha =c$) and $\pm 1/2$ HL spinons ($\alpha
=s$) by $L_{\alpha,\,\pm 1/2}$. As mentioned above, besides
corresponding to well defined occupancies of the BA quantum numbers,
the holons, spinons, and pseudofermions can also be expressed in terms
of rotated electrons. For instance, $N_{c0}$ equals the number of
rotated-electron singly occupied sites and $[N_a -N_{c0}]$ equals the
number of rotated-electron doubly occupied plus unoccupied sites. We
call $M_{\alpha,\,\pm 1/2}$ the number of $\pm 1/2$ holons ($\alpha
=c$) and $\pm 1/2$ spinons ($\alpha =s$). The latter number and that
of $\pm 1/2$ Yang holons ($\alpha =c$) and $\pm 1/2$ HL spinons
($\alpha =s$) are given by $M_{\alpha,\,\pm 1/2}=L_{\alpha,\,\pm 1/2}
+\sum_{\nu =1}^{\infty}\nu\,N_{\alpha\nu}$ and $L_{\alpha,\,\pm
1/2}=S_{\alpha} \mp S_{\alpha}^z$, respectively.

Within the pseudofermion, Yang holon, and HL spinon description
the energy and momentum spectrum of the PS energy eigenstates 
has the form provided in Eqs. (28)-(34) of Ref. \cite{V-1}. 
Such a spectrum is expressed in
terms of the pseudofermion energy dispersions defined in
Eqs. (C.15)-(C.18) of Ref. \cite{I}, pseudofermion bare-momentum
distribution-function deviations given in Eqs.  (13)-(17) of
Ref. \cite{V-1}, and Yang holon ($\alpha =c$) and HL spinon ($\alpha
=s$) occupancies $L_{\alpha,\,\pm 1/2}=S_{\alpha} \mp S_{\alpha}^z$.
(The pseudofermion energy dispersions equal those 
plotted in Figs. 6-9 of Ref. \cite{II}.)  As explained
in detail in Ref. \cite{I}, the number of $\pm 1/2$ holons and $\pm
1/2$ spinons can be expressed in terms of the number $N_{\sigma}$ of
spin-projection $\sigma$ electrons and rotated electrons, $N^h= [2N_a -N]$ 
of electronic holes and rotated-electron holes, and $N_{c0}$ of 
rotated-electron singly occupied sites as $M_{c,\,-1/2} = [N-N_{c0}]/2$, 
$M_{c,\,+1/2}=[N^h-N_{c0}]/2$, $M_{s,\,-1/2} = [N_{c0}-N_{\uparrow}
+N_{\downarrow}]/2$, and $M_{s,\,+1/2}=[N_{c0}+N_{\uparrow}-
N_{\downarrow}]/2$.  We recall that the number $N_{c0}$ of 
rotated-electron singly occupied sites also equals the number of 
$c0$ pseudofermions and the number $[N_a -N_{c0}]$ of
rotated-electron doubly-occupied and unoccupied sites equals
that of $c0$ pseudofermion holes. Furthermore, 
$M_{\alpha}=[M_{\alpha,\,-1/2}+M_{\alpha,\,+1/2}]$ 
denotes the number of holons ($\alpha =c$) or spinons ($\alpha =s$) 
such that $M_c =[N_a-N_{c0}]$ and $M_s =N_{c0}$ and
$L_{\alpha}=[L_{\alpha,\,-1/2}+L_{\alpha,\,+1/2}]$ denotes the number
of Yang holons ($\alpha =c$) or HL spinons ($\alpha =s$) such that
$L_c =2S_c = 2\eta$ and $L_s =2S_s = 2S$. Often in this paper we use
the notation $\alpha\nu\neq c0,\,s1$ branches, which refers to all
$\alpha\nu$ branches except the $c0$ and $s1$ branches. Moreover, the
summation (and product) $\sum_{\alpha\nu}$ (and $\prod_{\alpha\nu}$)
runs over all $\alpha\nu$ branches with finite $\alpha\nu$
pseudofermion occupancy in the corresponding state or subspace
and the summation $\sum_{\alpha}$ runs over $\alpha =c,\,s$.  An
important point for our studies is that for a ground state with
densities in the ranges considered in this paper the above numbers
read $N_{c0}=N$, $N_{s1}= N_{\downarrow}$,
$M_{c,\,+1/2}=L_{c,\,+1/2}=[N_a -N]$, $M_{s,\,-1/2}=N_{\downarrow}$,
$M_{s,\,+1/2}=N_{\uparrow}$,
$L_{s,\,+1/2}=[N_{\uparrow}-N_{\downarrow}]$, and
$N_{\alpha\nu}=M_{c,\,-1/2}=L_{\alpha,\,-1/2}=0$ for $\alpha\nu\neq
c0,\,s1$ and $\alpha =c,\,s$.

The $\alpha\nu$ pseudofermion discrete canonical-momentum values ${\bar{q}}_j$ are of the
following form,
\begin{equation}
{\bar{q}}_j = {\bar{q}} (q_j) = q_j + {Q^{\Phi}_{\alpha\nu} (q_j)\over L} = {2\pi\over
L}I^{\alpha\nu}_j + {Q^{\Phi}_{\alpha\nu} (q_j)\over L} \, ; \hspace{0.5cm}
j=1,2,...,N_{\alpha\nu}^* \, . \label{barqan}
\end{equation}
Here $I^{\alpha\nu}_j$ are the actual quantum numbers which are integers or half-odd
integers \cite{I} and the discrete bare-momentum $q_j$ such that $q_{j+1}-q_j =2\pi/L$
has allowed occupancies one and zero only. The corresponding discrete canonical-momentum 
${\bar{q}}_j$ such that ${\bar{q}}_{j+1}-{\bar{q}}_j =2\pi/L + {\cal O}(1/L^2)$ has
also allowed occupancies one and zero only. Thus, the bare-momentum distribution
function $N_{\alpha\nu} (q_j)$ is such that $N_{\alpha\nu} (q_j)=1$ for occupied
bare-momentum values and $N_{\alpha\nu} (q_j)=0$ for unoccupied bare-momentum values. 
We denote the ground-state bare-momentum distribution function by $N^{0}_{\alpha\nu} (q_j)$.
It is given in Eqs. (C.1)-(C.3) of Ref. \cite{I}. Except for $1/L$ corrections, for initial ground states
with densities in the ranges considered here the $c0$ and $s1$ pseudofermion 
{\it Fermi} momentum values $\pm q^0_{F\alpha\nu}$ and limiting bare-momentum values 
$\pm q^0_{\alpha\nu}$ of the $\alpha\nu$ band are such that,
\begin{equation}
q^0_{Fc0} = 2k_F \, ; \hspace{0.35cm} q^0_{Fs1} = k_{F\downarrow} \, ; \hspace{0.35cm}
q^0_{c0} = \pi \, ; \hspace{0.35cm} q^0_{s1} = k_{F\uparrow} \, ; \hspace{0.35cm}
q^0_{c\nu} = [\pi -2k_F] \, , \hspace{0.15cm} \nu
>0 \, ; \hspace{0.35cm} q^0_{s\nu} =
[k_{F\uparrow}-k_{F\downarrow}] \, , \hspace{0.15cm} \nu >1
\, . \label{q0Fcs}
\end{equation}
For the PS where the pseudofermion representation is defined, the set
of limiting values given in Eq. (\ref{q0Fcs}) also gives the corresponding
canonical-momentum limiting values, which remain unchanged for the 
excited states \cite{S-P}. (The ground-state {\it Fermi} bare-momentum 
values and limiting bare-momentum values including the $1/L$ corrections 
are given in Eqs. (C.4)-(C.11) and (B.14)-(B.17), respectively, of Ref. \cite{I}.)  

A $\alpha\nu$ pseudofermion can be labeled by the bare-momentum $q_j$
or corresponding canonical momentum ${\bar{q}}_j$. Indeed, there is a one-to-one
correspondence between the bare momentum $q_j$ and the canonical momentum
${\bar{q}}_j = q_j + Q^{\Phi}_{\alpha\nu} (q_j)/L$ for $j=1,2,...,N_{\alpha\nu}^*$. 
A $\alpha\nu$ pseudofermion (and $\alpha\nu$ pseudofermion hole) of 
bare-momentum $q_j$ corresponds to an occupied (and unoccupied)
BA quantum number $I^{\alpha\nu}_j$ of Eq. (\ref{barqan}). The above
number $N^*_{\alpha\nu}$ is such that $N^*_{\alpha\nu}=N_{\alpha\nu}+N_{\alpha\nu}^h$ 
where $N_{\alpha\nu}^h$ denotes the number of $\alpha\nu$ pseudofermion holes. 
(The expression of $N^h_{\alpha\nu}$ is given in Eqs.
(B.7) and (B.8) of Ref. \cite{I}.) Note that besides equaling the number of discrete
canonical-momentum values in the $\alpha\nu$ band,
$N^*_{\alpha\nu}$ also equals the number of sites of the
$\alpha\nu$ effective lattice \cite{IIIb}, which plays an important role in the
pseudofermion description. In addition to the $\alpha\nu$ pseudofermions of canonical
momentum ${\bar{q}}_j$, there are local $\alpha\nu$ pseudofermions, whose creation and
annihilation operators correspond to the sites of the effective $\alpha\nu$ lattice. Such
a lattice has spatial coordinates $x_j =a_{\alpha\nu}\,j$ where $a_{\alpha\nu} = 
L/N^*_{\alpha\nu}$ is the effective $\alpha\nu$ lattice constant and
$j=1,2,...,N^*_{\alpha\nu}$. Each $\alpha\nu$ pseudofermion band is associated
with an effective $\alpha\nu$ lattice whose length $L=N^*_{\alpha\nu}\,a_{\alpha\nu}$ is
the same as that of the original real-space lattice. The canonical-momentum pseudofermion 
operators and local pseudofermion operators are related by a 
Fourier transform \cite{IIIb}.

The canonical-momentum shift functional $Q^{\Phi}_{\alpha\nu} (q_j)/L$ appearing 
in the canonical-momentum expression (\ref{barqan}) is given by,
\begin{equation}
{Q^{\Phi}_{\alpha\nu} (q_j)\over L} = {2\pi\over L} \sum_{\alpha'\nu'}\,\,
\sum_{j'=1}^{N^*_{\alpha'\nu'}}\,\Phi_{\alpha\nu,\,\alpha'\nu'}(q_j,q_{j'})\, \Delta
N_{\alpha'\nu'}(q_{j'}) \, , \label{qcan1j}
\end{equation}
where $\Delta N_{\alpha\nu} (q_j) \equiv N_{\alpha\nu} (q_j) - N^{0}_{\alpha\nu} (q_j)$
is the $\alpha\nu$ bare-momentum distribution-function deviation. Thus,
${\bar{q}}_j = q_j$ for the initial ground state. A PS excited energy
eigenstate is uniquely defined by the values of the set of deviations $\{\Delta
N_{\alpha\nu} (q_j)\}$ for all values of $q_j$ corresponding to the $\alpha\nu$
branches with finite pseudofermion occupancy in the state and by the values $L_{c
,\,-1/2}$ and $L_{s ,\,-1/2}$. The quantity
$\Phi_{\alpha\nu,\,\alpha'\nu'}(q,q')$ on the right-hand side of Eq. (\ref{qcan1j}) is a
function of both the bare-momentum values $q$ and $q'$ given by,
\begin{equation}
\Phi_{\alpha\nu,\,\alpha'\nu'}(q,q') = \bar{\Phi }_{\alpha\nu,\,\alpha'\nu'}
\left({4t\,\Lambda^{0}_{\alpha\nu}(q)\over U}, {4t\,\Lambda^{0}_{\alpha'\nu'}(q')\over
U}\right) \, , \label{Phi-barPhi}
\end{equation}
where the function $\bar{\Phi }_{\alpha\nu,\,\alpha'\nu'} (r ,\,r')$ is the unique
solution of the integral equations (A1)-(A13) of Ref. \cite{IIIb}. The ground-state
rapidity functions $\Lambda_{\alpha\nu}^0 (q)$ appearing in Eq. (\ref{Phi-barPhi}), where
$\Lambda^0_{c0}(q)\equiv\sin k^0 (q)$ for $\alpha\nu=c0$, are defined in terms of the
inverse functions of $k^0 (q)$ and $\Lambda_{\alpha\nu}^0 (q)$ for $\nu
>0$ in Eqs. (A.1) and (A.2) of Ref. \cite{V-1}. 
As discussed below, $2\pi\,\Phi_{\alpha\nu,\,\alpha'\nu'}(q,q')$ [or
$-2\pi\,\Phi_{\alpha\nu,\,\alpha'\nu'}(q,q')$] is an elementary {\it two-pseudofermion
phase shift} such that $q$ is the bare-momentum value of a $\alpha\nu$ pseudofermion or
$\alpha\nu$ pseudofermion hole scattered by a $\alpha'\nu'$ pseudofermion [or
$\alpha'\nu'$ pseudofermion hole] of bare-momentum $q'$ created under a ground-state -
excited-energy-eigenstate transition. For initial ground states with electronic 
density $n=1$ (and spin density $m=0$) and $c\nu\neq c0$ or $c\nu'\neq c0$ 
branches (and $s\nu\neq s1$ or $s\nu'\neq s1$ branches), the ground-state 
rapidity function  $\Lambda^{0}_{\alpha\nu}(q)$ or $\Lambda^{0}_{\alpha'\nu'}(q')$ 
appearing in expression (\ref{Phi-barPhi}) must be 
replaced by that of the excited state described by the bare-momentum 
distribution-function deviations on the right-hand site of Eq. (\ref{qcan1j})
\cite{S-P}.

%%%%%%%%%%%%%%%%%%%%%%%%%%%%%%%%%%%%%%%%%%%%%%%%%%%%%%%%%%%%%%%%
\section{S MATRICES AND PHASE SHIFTS}

Here we consider the general pseudofermion and hole $S$ matrices and
phase shifts. Moreover, we relate the phase shifts and $S$ matrices of the 
two representations mentioned in Sec. I for the reduced subspace
considered in the studies of Refs. \cite{S0,S}.

%%%%%%%%%%%%%%%%%%%%%%%%%%%%%%%%%%%%%%%%%%%%%%%%%%%%%%%%%%%%%%%%
\subsection{PSEUDOFERMION AND HOLE S MATRICES}

Our analysis refers to periodic boundary conditions and very large values of $L$.
The PS energy and momentum eigenstates can be written as direct products of 
states spanned by the occupancy configurations of each of the $\alpha\nu$ 
branches with finite pseudofermion occupancy in the state under consideration. 
Moreover, the many-pseudofermion states spanned by occupancy configurations 
of each $\alpha\nu$ branch can be expressed as a direct product of 
$N^*_{\alpha\nu}$ one-pseudofermion states, each referring to one 
discrete bare-momentum value $q_j$, where $j=1,2,...,N^*_{\alpha\nu}$. 
Within the pseudofermion description, the 1D Hubbard model in normal order 
relative to the initial ground state reads \cite{IIIb} $:\hat{H}: =
 \sum_{\alpha\nu}\sum_{j=1}^{N^*_{\alpha\nu}}\hat{H}_{\alpha\nu,q_j} +
\sum_{\alpha}\hat{H}_{\alpha}$, where $\hat{H}_{\alpha\nu,q_j}$ is the 
one-pseudofermion Hamiltonian which describes the $\alpha\nu$ pseudofermion 
or hole of bare-momentum $q_j$ and $\hat{H}_{\alpha}$ refers to the 
Yang holons ($\alpha =c$) and HL spinons 
($\alpha =s$). (As discussed below, the latter objects are scatter-less.) 
For each many-pseudofermion PS energy eigenstate the number of Hamiltonians
$\hat{H}_{\alpha\nu,q_j}$ equals that of one-pseudofermion states given by,
$N^*_{c0} + N^*_{s1} + \sum_{\alpha\nu\neq c0 ,\,s1} \theta (\vert\Delta
N_{\alpha\nu}\vert)\, N^*_{\alpha\nu}$ where $\theta (x)=1$ for $x>0$ and 
$\theta (x)=0$ for $x= 0$ and the pseudofermion numbers refer to
the energy eigenstate under consideration. 

The ground-state - excited-energy-eigenstate transitions can be divided into
three steps. The first step refers to the ground-state - virtual-state
transition. It is scatter-less and changes the number 
of discrete bare-momentum values of the $\alpha\nu\neq c0$ bands.
Moreover, the first step involves the pseudofermion creation and annihilation processes and 
pseudofermion particle-hole processes associated with PS excited states.
The second step is also scatter-less and generates the "in" state. 
Indeed, the one-pseudofermion states belonging 
to the many-pseudofermion "in" state are the "in" asymptote states of the 
pseudofermion scattering theory. The generator of the virtual-state - "in"-state 
transition is of the form ${\hat{S}}^{0} = \prod_{\alpha\nu}\prod_{j=1}^{N^*_{\alpha\nu}}
{\hat{S}}^{0}_{\alpha\nu ,q_j}$ where ${\hat{S}}^{0}_{\alpha\nu ,q_j}$ is a 
well-defined one-pseudofermion unitary operator. Application of ${\hat{S}}^{0}_{\alpha\nu ,q_j}$ 
onto the corresponding one-pseudofermion state of the many-pseudofermion virtual 
state shifts its discrete bare-momentum value $q_j$ to the bare-momentum 
value $q_j+Q_{\alpha\nu}^0/L$, where $Q_{\alpha\nu}^0$ is given in 
Eq. (\ref{pic0an}) of Appendix A. Finally, the third step consists of a set of two-pseudofermion
scattering events. It corresponds to the "in"-state - "out"-state transition,
where the latter state is the PS excited energy eigenstate under
consideration. The generator of that transition is the operator,
${\hat{S}}^{\phi} = \prod_{\alpha\nu}\prod_{j=1}^{N^*_{\alpha\nu}}
{\hat{S}}^{\phi}_{\alpha\nu ,q_j}$ where ${\hat{S}}^{\phi}_{\alpha\nu ,q_j}$ is a 
well-defined one-pseudofermion scattering 
unitary operator. The one-pseudofermion states belonging to the 
many-pseudofermion "out" state are the "out" asymptote pseudofermion
scattering states. Application of ${\hat{S}}^{\phi}_{\alpha\nu ,q_j}$ onto the 
corresponding one-pseudofermion state of the many-pseudofermion "in" 
state shifts its discrete bare-momentum value $q_j+Q_{\alpha\nu}^0/L$ to the
"out"-state discrete canonical-momentum value $q_j+Q_{\alpha\nu} (q_j)/L$
where,
\begin{equation}
Q_{\alpha\nu}(q_j) = Q_{\alpha\nu}^0 + Q^{\Phi}_{\alpha\nu} (q_j) \, . \label{Qcan1j}
\end{equation}
We note that the generator of the virtual-state - "out"-state transition is 
the unitary operator ${\hat{S}}\equiv {\hat{S}}^{\phi}{\hat{S}}^{0} = 
\prod_{\alpha\nu}\prod_{j=1}^{N^*_{\alpha\nu}}{\hat{S}}_{\alpha\nu ,q_j}$ 
where ${\hat{S}}_{\alpha\nu ,q_j}$ is the one-pseudofermion or hole unitary 
${\hat{S}}_{\alpha\nu ,q_j}={\hat{S}}^{\phi}_{\alpha\nu ,q_j}{\hat{S}}^{0}_{\alpha\nu ,q_j}$ 
operator. The unitary ${\hat{S}}_{\alpha\nu ,q_j}$ operator 
shifts the discrete bare-momentum value $q_j$ of the one-pseudofermion
state belonging to the virtual state directly to the "out"-state discrete 
canonical-momentum value $q_j+Q_{\alpha\nu} (q_j)/L$.

The virtual state, "in" state, and "out" state are PS energy eigenstates, as 
further discussed below. Thus, that the one-pseudofermion states of the 
many-pseudofermion "in" state and "out" state are the one-pseudofermion 
"in" and "out" asymptote scattering states, respectively, implies that the 
one-pseudofermion Hamiltonian $\hat{H}_{\alpha\nu,q_j}$ plays the role 
of the unperturbed Hamiltonian $\hat{H}_0$ of the 
spin-less one-particle nonrelativistic scattering theory. Indeed, 
the unitary ${\hat{S}}_{\alpha\nu ,q_j}$ operator (and
the scattering unitary ${\hat{S}}^{\phi}_{\alpha\nu ,q_j}$ operator) commutes with the 
Hamiltonian $\hat{H}_{\alpha\nu,q_j}$ and thus the one-pseudofermion 
"in" and "out" asymptote scattering states are energy eigenstates of $\hat{H}_{\alpha\nu,q_j}$
and eigenstates of ${\hat{S}}_{\alpha\nu ,q_j}$ (and ${\hat{S}}^{\phi}_{\alpha\nu ,q_j}$). 
It follows that the matrix elements between one-pseudofermion states
of  ${\hat{S}}_{\alpha\nu ,q_j}$ (and ${\hat{S}}^{\phi}_{\alpha\nu ,q_j}$) are
diagonal and thus these operators are fully defined by the set of their eigenvalues
belonging to these states. The same applies to the above generator ${\hat{S}}$ 
(and ${\hat{S}}^{\phi}$). The matrix elements of that generator between 
virtual states (and "in" states) are also diagonal and thus it is fully defined 
by the set of its eigenvalues belonging to the virtual states (and "in" states). 
Since ${\hat{S}}^{\phi}_{\alpha\nu ,q_j}$ and ${\hat{S}}_{\alpha\nu ,q_j}$ are unitary, 
each of their eigenvalues has modulus one and can be written as the exponent of a 
purely imaginary number given by,
\begin{eqnarray}
S^{\Phi}_{\alpha\nu} (q_j) & = & e^{iQ^{\Phi}_{\alpha\nu}(q_j)} =
\prod_{\alpha'\nu'}\,\prod_{j'=1}^{N^*_{\alpha'\nu'}}\,S_{\alpha\nu ,\,\alpha'\nu'} (q_j, q_{j'}) \, ;
\hspace{0.25cm} j=1,2,..., N^*_{\alpha\nu} \nonumber \\
S_{\alpha\nu} (q_j) & = & e^{iQ_{\alpha\nu}(q_j)} =
e^{i\,Q_{\alpha\nu}^0}  S^{\Phi}_{\alpha\nu} (q_j) \, ;
\hspace{0.25cm} j=1,2,..., N^*_{\alpha\nu}
\, . \label{San}
\end{eqnarray}
Here $Q^{\Phi}_{\alpha\nu}(q_j)$ and $Q_{\alpha\nu}(q_j)$ are the functionals 
defined in Eqs. (\ref{qcan1j}) and (\ref{Qcan1j}), respectively.
By use of the former functional we find that,
\begin{equation}
S_{\alpha\nu ,\,\alpha'\nu'} (q_j, q_{j'}) =
e^{i2\pi\,\Phi_{\alpha\nu,\,\alpha'\nu'}(q_j,q_{j'})\, \Delta N_{\alpha'\nu'}(q_{j'})}
\, , \label{Sanan}
\end{equation}
where the functions $\pi\,\Phi_{\alpha\nu,\,\alpha'\nu'}(q_j,q_{j'})$ are
given in Eq. (\ref{Phi-barPhi}). The effect of 
under a ground-state - excited-energy-eigenstate transition moving the 
$\alpha\nu$ pseudofermion or hole of initial ground-state canonical-momentum 
${\bar{q}}_j=q_j$ once around the length $L$ lattice ring is that its wave function acquires 
the overall phase factor $S_{\alpha\nu} (q_j)$ given in Eq. (\ref{San}).
The phase factor $S_{\alpha\nu ,\,\alpha'\nu'} (q_j, q_{j'})$ of Eq.  (\ref{Sanan}) 
in the wave function of the $\alpha\nu$ pseudofermion or hole results from a elementary 
two-pseudofermion zero-momentum forward-scattering event whose scattering
center is a $\alpha'\nu'$ pseudofermion ($\Delta N_{\alpha'\nu'}(q_{j'})=1$)
or $\alpha'\nu'$ pseudofermion hole ($\Delta N_{\alpha'\nu'}(q_{j'})=-1$) created
under the ground-state - excited-state transition. Thus, the third step 
of that transition involves a well-defined set of elementary two-pseudofermion 
scattering events where all $\alpha\nu$ pseudofermions and $\alpha\nu$ 
pseudofermion holes of bare-momentum $q_j+Q_{\alpha\nu}^0/L$ of the "in" 
state are the scatterers, which leads to the overall scattering phase factor 
$S^{\Phi}_{\alpha\nu} (q_j)$ in their wave function provided in Eq. (\ref{San}). 
That the scattering centers are the $\alpha'\nu'$ pseudofermions or 
pseudofermion holes of momentum $q_{j'}+Q_{\alpha\nu}^0/L$ created 
under the ground-state - excited-energy-eigenstate transition is confirmed 
by noting that $S_{\alpha\nu ,\,\alpha'\nu'} (q_j,q_{j'})=1$ for 
$\Delta N_{\alpha'\nu'}(q_{j'}) =0$. Thus, out of the scatterers whose 
number equals that of the one-pseudofermion states
given above, the scattering centers are only those whose 
bare-momentum distribution-function deviation is finite. The elementary 
two-pseudofermion scattering processes associated with the phase
factors (\ref{Sanan}) conserve the total energy and total momentum, 
are of zero-momentum forward-scattering type and thus conserve the 
individual "in" asymptote $\alpha\nu$ pseudofermion momentum value 
$q_j+Q_{\alpha\nu}^0/L$ and energy, and also conserve the $\alpha\nu$ branch, 
usually called {\it channel} in the scattering language.
Moreover, the scattering amplitude does not connect quantum objects with 
different $\eta$ spin or spin.

Importantly, for each $\alpha\nu$ pseudofermion or pseudofermion hole of 
virtual-state bare-momentum $q_j$, the $S$ matrix associated with the 
ground-state - excited-energy-eigenstate transition is simply the phase 
factor $S_{\alpha\nu} (q_j)$ given in Eq. (\ref{San}).
Application of the unitary ${\hat{S}}_{\alpha\nu ,q_j}$ operator onto
its one-pseudofermion state of the many-pseudofermion 
virtual state generates the corresponding one-pseudofermion 
state of the many-pseudofermion "out" state. The latter one-pseudofermion 
state equals the former one multiplied by the phase factor 
$S_{\alpha\nu} (q_j)$ of Eq. (\ref{San}).
(Applying the scattering unitary ${\hat{S}}^{\Phi}_{\alpha\nu ,q_j}$ operator 
onto its one-pseudofermion state of the many-pseudofermion
"in" state also generates the corresponding one-pseudofermion 
state of the many-pseudofermion "out" state; The latter one-pseudofermion 
state equals the former one multiplied by the phase factor 
$S^{\Phi}_{\alpha\nu} (q_j)$ of Eq. (\ref{San}).) It follows that the 
many-pseudofermion virtual states (and "in" states) are eigenstates
of the above generator ${\hat{S}}$ (and ${\hat{S}}^{\phi}$). The eigenvalue $S_T$
of ${\hat{S}}$ belonging to a PS virtual state and the eigenvalue 
$S^{\Phi}_T$ of ${\hat{S}}^{\Phi}$ belonging to a PS "in" state
are given by,
\begin{eqnarray}
S_T & = & e^{iQ_T} = \prod_{\alpha\nu}\prod_{j=1}^{N^*_{\alpha\nu}}
S_{\alpha\nu} (q_j) \, ; \hspace{0.5cm}
Q_T = \sum_{\alpha\nu}\sum_{j=1}^{N^*_{\alpha\nu}} 
Q_{\alpha\nu}(q_j) \nonumber \\
S^{\Phi}_T & = & e^{iQ^{\Phi}_T} = \prod_{\alpha\nu}\prod_{j=1}^{N^*_{\alpha\nu}}
S^{\Phi}_{\alpha\nu} (q_j) \, ; \hspace{0.5cm}
Q^{\Phi}_T = \sum_{\alpha\nu}\sum_{j=1}^{N^*_{\alpha\nu}} 
Q^{\Phi}_{\alpha\nu}(q_j) \, .
\label{ST}
\end{eqnarray}
The "out" state equals the virtual state multiplied by the
phase factor $S_T$ and the "in" state multiplied by $S^{\Phi}_T$. 
Since the "out" state is by construction an energy 
eigenstate of the 1D Hubbard model, this result confirms that the corresponding 
virtual and "in" states are also energy eigenstates of the model. The general expressions 
(\ref{qcan1j}) and (\ref{Qcan1j}) for the functionals $Q_{\alpha\nu}^{\Phi}(q_j)$ 
and $Q_{\alpha\nu} (q_j)$ define uniquely the eigenvalues $S^{\Phi}_T$
and $S_T$ of ${\hat{S}}^{\phi}$ and $\hat{S}$ for any PS "in" state and virtual 
state, respectively. Since these many-pseudofermion states equal the "out" 
excited energy eigenstate except for a phase factor, in this paper we often 
associate both $S$ matrices $S_{\alpha\nu} (q_j)$ and $S^{\Phi}_{\alpha\nu} (q_j)$
of Eq. (\ref{San}) indifferently with the corresponding excited energy eigenstate,
yet they are eigenvalues of one-pseudofermion states of the virtual and "in"
states, respectively.

When moving around the lattice ring the $\alpha\nu$ pseudofermion (or hole)
departures from the point $x=0$ and arrives to $x=L$, one finds that,
\begin{equation}
\lim_{x\rightarrow L}\,\bar{q}\,x = q\,x +Q_{\alpha\nu}^0 + Q^{\Phi}_{\alpha\nu} (q) =
q\,x + Q_{\alpha\nu} (q) \, , \label{qr1}
\end{equation}
where $q$ refers to the virtual state. For this asymptote coordinate choice,
$Q_{\alpha\nu} (q)$ is the overall $\alpha\nu$ pseudofermion (or hole) phase shift 
whose value is defined only to within addition of an arbitrary multiple of $2\pi$.
From analysis of Eqs. (\ref{qcan1j}) and (\ref{Qcan1j}) it follows that
$2\pi\,\Phi_{\alpha\nu,\,\alpha'\nu'}(q_j,q_{j'})$ is an elementary two-pseudofermion
phase shift. The studies of Refs. \cite{S-P0,S-P} consider other asymptote coordinates 
usually used in standard quantum non-relativistic scattering theory, 
such that $x\in (-L/2,\,+L/2)$ and thus $\delta_{\alpha\nu} (q) = 
Q_{\alpha\nu} (q)/2$ is the overall $\alpha\nu$ pseudofermion or hole phase shift 
given only to within addition of an arbitrary multiple of $\pi$. Furthermore,
$\pi\,\Phi_{\alpha\nu,\,\alpha'\nu'}(q_j,q_{j'})$ is an elementary two-pseudofermion 
phase shift. However, the choice of either definition is a matter of taste and the
uniquely defined quantity is the $S$ matrix.

Finally, an important property of the pseudofermion scattering 
theory introduced in Refs. \cite{S-P0,S-P} is
that the $\pm 1/2$ Yang holons and $\pm 1/2$ HL spinons are scatter-less objects.
Indeed, the form of the scattering part of the overall phase shift (\ref{Qcan1j}), Eq.
(\ref{qcan1j}), reveals that the value of such a phase-shift functional is independent of
the changes in the occupation numbers of the $\pm 1/2$ Yang holons and $\pm 1/2$ HL
spinons. Thus, these objects are not scattering centers. Moreover, they are not
scatterers, once their wave functions do not acquire any phase factor under the 
"in"-state - "out"-state transitions.

%%%%%%%%%%%%%%%%%%%%%%%%%%%%%%%%%%%%%%%%%%%%%%%%%%%%%%%%%%%%%%%%
\subsection{PHASE SHIFTS IN THE REDUCED SUBSPACE}
 
Within the pseudofermion description, the $m=0$ and $n=1$ initial
ground state is such that the $c0$ and $s1$ bands are full and thus
the {\it Fermi} momentum values given in Eq. (\ref{q0Fcs}) coincide
with the corresponding limiting values provided in the same equation
and are such that $q^0_{Fc0}=q^0_{c0}=\pi$ and
$q^0_{Fs1}=q^0_{s1}=\pi/2$.  The reduced subspace considered in the
studies of Refs. \cite{Natan, S0,S} is spanned by twelve types of
excited energy eigenstates of that ground state. The excited states
belonging to each of these types are characterized by fixed values for
the numbers of holes in the $c0$ and $s1$ bands, $\eta$ spin
$\eta=S_c=\Delta S_c$, $\eta$-spin projection $\eta_z=S_c^z=\Delta
S_c^z$, spin $S=S_s=\Delta S_s$, and spin projection $S_z=S_s^z=\Delta
S_s^z$. Here $\Delta S_c$, $\Delta S_c^z$, $\Delta S_s$, and $\Delta
S_s^z$ are the deviations relative to the values of the $m=0$ and
$n=1$ initial ground state. For excited states of that state such
deviations equal the corresponding values of 
$S_c$, $S_c^z$, $S_s$, and $S_s^z$,
respectively. If one specifies the two bare-momentum values of the
created holes, each of such classes of states corresponds to a
uniquely defined excited energy eigenstate. Thus, each class of states
is generated from one of these energy eigenstates by considering all
possible bare-momentum values of the two created pseudofermion holes.

The values of the numbers which characterize each class of excited states are
provided in Table I. In order to give information about the pseudofermions,
Yang holons, and HL spinons  created or annihilated under the 
ground-state - excited-state transitions, we also provide in that
table the values for the deviations $\Delta N_{c0}$, $\Delta N_{s1}$, 
$N_{c1}=\Delta N_{c1}$, $N_{s2}=\Delta N_{s2}$, $L_{c,\,-1/2} =\Delta L_{c,\,-1/2}$, 
$\Delta L_{c,\,+1/2}$, $L_{s,\,-1/2} =\Delta L_{s,\,-1/2}$, and $\Delta L_{s,\,+1/2}$.  
Indeed, a PS {\it CPHS ensemble subspace} plays an important role in the 
pseudofermion representation \cite{V-1,LE}
and is spanned by all energy eigenstates with the same values for the 
sets of numbers $N_{c0}$, $\{N_{c\nu}\}$, $\{N_{s\nu}\}$, $\{L_{c,\,-1/2}\}$, and 
$\{L_{s,\,-1/2}\}$ such that $\nu =1,2,...$.
In turn, an {\it electronic ensemble space} is spanned by all energy 
eigenstates with the same values for the electronic numbers $N_{\uparrow}$ and 
$N_{\downarrow}$ and a {\it CPHS ensemble space} is spanned by 
all energy eigenstates with the same values for the numbers $\{M_{\alpha,\,\pm 1/2}\}$ 
of $\pm 1/2$ holons ($\alpha =c$) and $\pm 1/2$ spinons ($\alpha =c$) \cite{S-P}. 
(In CPHS ensemble space, CPHS refers to $c0$ pseudofermion, holon, and spinon.) 
In Table II we provide the values of the corresponding deviations $\Delta N_{\sigma}$ 
in the spin-projection $\sigma$ electronic numbers and  $\Delta M_{\alpha,\,\pm 1/2}$ 
in the $\pm 1/2$ holon ($\alpha =c$) and $\pm 1/2$ spinon ($\alpha =s$) numbers of 
the pseudofermion representation of Refs. \cite{I,IIIb} for each class of excited states of Table I. 
Moreover, in Table II we also provide the values of the scatter-less phase shifts  
$Q^0_{c0}$ and $Q^0_{s1}$ given in Eq. (\ref{pic0an}) of Appendix A for these states. We 
note that the numbers of spinons and holons of the alternative spinon-holon 
representation of Refs. \cite{Natan,S0,S} equal the numbers $N_{s1}^h =\Delta N_{s1}^h$ 
of spin holes and $N_{c0}^h=\Delta N_{c0}^h$ of charge holes, respectively, 
given in Table I.

\begin{table}
\begin{tabular}{|c||c|c|c|c|c|c|c|c|c|c|c|c|c|c|} \hline
Excited state & Charge holes & Spin holes & $\eta$ & $\eta_z$ & $S$ & $S_z$ & 
$\Delta N_{c0}$ & $\Delta N_{s1}$ & $N_{c1}$ & $N_{s2}$ & 
$L_{c,\,-1/2}$ & $\Delta L_{c,\,+1/2}$ & $L_{s,\,-1/2}$ & $\Delta L_{s,\,+1/2}$ \\
\hline\hline
$\eta$-spin triplet & 2 & 0 & 1 & 1 & 0 & 0 & -2 & -1 & 0 & 0 & 0 & 2 & 0 & 0 \\
\hline
$\eta$-spin triplet & 2 & 0 & 1 & 0 & 0 & 0 & -2 & -1 & 0 & 0 & 1 & 1 & 0 & 0 \\
\hline
$\eta$-spin triplet & 2 & 0 & 1 & -1 & 0 & 0 & -2 & -1 & 0 & 0 & 2 & 0 & 0 & 0 \\
\hline
$\eta$-spin singlet & 2 & 0 & 0 & 0 & 0 & 0 & -2 & -1 & 1 & 0 & 0 & 0 & 0 & 0 \\
\hline
spin triplet & 0 & 2 & 0 & 0 & 1 & 1 & 0 & -1 & 0 & 0 & 0 & 0 & 0 & 2 \\
\hline
spin triplet & 0 & 2 & 0 & 0 & 1 & 0 & 0 & -1 & 0 & 0 & 0 & 0 & 1 & 1 \\
\hline
spin triplet & 0 & 2 & 0 & 0 & 1 & -1 & 0 & -1 & 0 & 0 & 0 & 0 & 2 & 0 \\
\hline
spin singlet & 0 & 2 & 0 & 0 & 0 & 0 & 0 & -2 & 0 & 1 & 0 & 0 & 0 & 0 \\
\hline
doublet & 1 & 1 & 1/2 & 1/2 & 1/2 & 1/2 & -1 & -1 & 0 & 0 & 1 & 0 & 1 & 0 \\
\hline
doublet & 1 & 1 & 1/2 & -1/2 & 1/2 & 1/2 & -1 & -1 & 0 & 0 & 0 & 1 & 1 & 0 \\
\hline
doublet & 1 & 1 & 1/2 & 1/2 & 1/2 & -1/2 & -1 & -1 & 0 & 0 & 1 & 0 & 0 & 1 \\
\hline
doublet & 1 & 1 & 1/2 & -1/2 & 1/2 & -1/2 & -1 & -1 & 0 & 0 & 0 & 1 & 0 & 1 \\
\hline
\end{tabular}
\caption{The twelve types of energy eigenstates that span the
reduced subspace. The number of charge and spin holes refers
to the numbers $N^h_{c0}=\Delta N^h_{c0}$ and $N^h_{s1}=\Delta N^h_{s1}$, 
respectively. For fixed and different momentum values of the two holes, each 
state type corresponds to a well defined excited energy eigenstate 
of the $n=1$ and $m=0$ initial ground state. Since the symmetry of
the problem is $SO(4)$ rather than $SU(2)\times SU(2)$, note that
only energy eigenstates such that $\eta +S$ is an integer number
are allowed. The reduced subspace is spanned by all two-hole and
$\eta +S=1$ excited states of the $\eta +S=0$ ground state.}
\end{table}

Within the pseudofermion description, two out of the twelve types of excited states
of Table I have one $\alpha\nu\neq c0,\,s1$ pseudofermion: the $\eta$-spin singlet 
excited states and spin singlet excited states have one $c1$ pseudofermion and one $s2$ 
pseudofermion, respectively. While for initial ground states with densities in the 
ranges $0<n<1$ and $0<m<n$ the $\alpha\nu\neq c0,\,s1$ 
pseudofermions are scatterers, it was 
found in Ref. \cite{S-P} that for excited energy eigenstates of a $n=1$ (and $m=0$)
initial ground state with a single $c\nu\neq c0$ pseudofermion (and $s\nu\neq s1$ 
pseudofermion), such a quantum object has bare momentum $q=0$, canonical momentum 
$\bar{q}=q=0$, and is invariant under the electron - rotated-electron 
unitary transformation. Since $\bar{q}=q=0$, such a $c\nu\neq c0$ 
pseudofermion (and $s\nu\neq s1$ pseudofermion) is not a
scatterer. This implies that $Q_{c1} (0)=0$ and $Q_{s2} (0)=0$ for the overall 
phase shift given in Eq. (\ref{Qcan1j})  
corresponding to the $c1$ pseudofermion and $s2$ pseudofermion,
respectively, created under the transition from the ground state 
to the $\eta$-spin singlet and spin singlet excited state, respectively.
It follows that for the $n=1$ (and $m=0$) initial ground state there
is no one-pseudofermion scattering state for the 
$c1$ pseudofermion (and $s2$ pseudofermion). Thus, 
within the pseudofermion scattering theory, for all the reduced-subspace excited 
energy eigenstates of Table I the only quantum objects that are both
scatterers and scattering centers are the $c0$ pseudofermion 
holes and/or the $s1$ pseudofermion holes created under the 
ground-state - excited-state transitions. 

Here we calculate all the $c0$ and $s1$ pseudofermion-hole overall phase shifts 
associated with the types of excited states of Table I. Interestingly, we show that 
for such excited states the overall pseudofermion-hole phase shifts $Q_{c0} (q)$ and 
$Q_{s1} (q)$ defined by the general overall phase-shift expression (\ref{Qcan1j}) 
have the same values as the holon and spinon phase shifts, respectively, considered 
in Refs. \cite{Natan,S0,S}. (For the phase shift $Q_{c0} (q)$ this is true except for a constant 
term, as further discussed below.) 

Let us show that the phase shifts provided in Eqs. (5.19)-(5.21) of Ref.
\cite{S} correspond indeed to particular cases of the overall pseudofermion-hole phase shift
functionals $Q_{c0} (q)$ and $Q_{s1} (q)$ defined by Eq. (\ref{Qcan1j}). (We recall
that such phase shifts refer to a pseudofermion hole when the corresponding
bare-momentum value $q$ is empty for the excited state.) In Appendix A
we provide the bare-momentum distribution-function deviations of the
twelve classes of excited states of Tables I and II. Use of 
Eqs. (\ref{DNc-eT})-(\ref{DNcs1-eD}) of that Appendix in Eqs. (\ref{qcan1j}) 
and (\ref{Qcan1j}) for the overall phase shift leads to,
\begin{equation}
Q_{c0} (q) =
-\pi\Bigl[2\sum_{l=1}^{2}\Phi_{c0,\,c0}(q,q_l)-\Phi_{c0,\,c0}(q,\pi)+\Phi_{c0,\,c0}(q,-\pi)+
\Phi_{c0,\,s1}(q,\pi/2)+\Phi_{c0,\,s1}(q,-\pi/2)\Bigr] \, , \label{Qc-eT}
\end{equation}
for the three classes of $\eta$-spin triplet states,
\begin{equation}
Q_{c0} (q) = -\pi\Bigl[2\sum_{l=1}^{2}\Phi_{c0,\,c0}(q,q_l)+
\Phi_{c0,\,s1}(q,\pi/2)+\Phi_{c0,\,s1}(q,-\pi/2)-2\Phi_{c0,\,c1}(q,0)\Bigr] \, ,
\label{Qc-eS}
\end{equation}
for the $\eta$-spin singlet states,
\begin{equation}
Q_{s1} (q) =
-\pi\Bigl[2\sum_{l=1}^{2}\Phi_{s1,\,s1}(q,{q'}_l)-\Phi_{s1,\,c0}(q,\pi)+\Phi_{s1,\,c0}(q,-\pi)-
\Phi_{c0,\,s1}(q,\pi/2)-\Phi_{c0,\,s1}(q,-\pi/2)\Bigr] \, , \label{Qs1-T}
\end{equation}
for the three classes of spin triplet states,
\begin{equation}
Q_{s1} (q) = -\pi\Bigl[2\sum_{l=1}^{2}\Phi_{s1,\,s1}(q,{q'}_l)-\Phi_{s1,\,c0}(q,\pi)
+\Phi_{s1,\,c0}(q,-\pi)-2\Phi_{s1,\,s2}(q,0)\Bigr] \, , \label{Qs1-S}
\end{equation}
for the spin singlet states and,
\begin{equation}
Q_{c0} (q) = -\pi\Bigl[2\Phi_{c0,\,c0}(q,q_1)+ 2\Phi_{c0,\,s1}(q,{q'}_1)\Bigr] \, ;
\hspace{1cm} Q_{s1} (q) = -\pi\Bigl[2\Phi_{s1,\,c0}(q,q_1)+
2\Phi_{s1,\,s1}(q,{q'}_1)\Bigr] \, , \label{Qc-s1-eD}
\end{equation}
for the four classes of doublet states.

By taking the limits $m\rightarrow 0$ and $n\rightarrow 1$ in the above expressions 
(\ref{Qc-eT})-(\ref{Qc-s1-eD}) for the phase shifts $Q_{c0} (q)$ and $Q_{s1} (q)$ 
at $q=q_1$ and $q={q'}_1$, respectively, we find,
\begin{equation}
Q_{c0} (q_1) = 2\pi B \Bigl({\sin k_1 -\sin k_2\over u}\Bigr) = \delta_{CT} - \pi \, , \label{Qc-1-eT}
\end{equation}
for the $\eta$-spin triplet states,
\begin{equation}
Q_{c0} (q_1) = -2\arctan \Bigl({\sin k_1 -\sin k_2\over 2u}\Bigr) +2\pi B \Bigl({\sin k_1
-\sin k_2\over u}\Bigr)  = \delta_{CS} -\pi \, , \label{Qc-1-eS}
\end{equation}
for the $\eta$-spin singlet states,
\begin{equation}
Q_{s1} ({q'}_1) =-2\pi B \Bigl({{\Lambda'}_1 -{\Lambda'}_2\over u}\Bigr) = \delta_{ST} \, ,
\label{Qs1-1-T}
\end{equation}
for the spin triplet states,
\begin{equation}
Q_{s1} ({q'}_1) =2\arctan \Bigl({{\Lambda'}_1 -{\Lambda'}_2\over 2u}\Bigr) -2\pi B
\Bigl({{\Lambda'}_1 -{\Lambda'}_2\over u}\Bigr) = \delta_{SS} \, , \label{Qs1-1-S}
\end{equation}
for the spin singlet states and,
\begin{equation}
Q_{c0} (q_1) = \arctan \Bigl(\sinh\Bigl({\pi\over 2}\Bigl[{\sin k_1 -{\Lambda'}_1\over
u}\Bigr]\Bigr)\Bigr) = \delta_{\eta S} - \pi \, ; \hspace{0.5cm} Q_{s1} ({q'}_1) = \arctan
\Bigl(\sinh\Bigl({\pi\over 2}\Bigl[{{\Lambda'}_1 -\sin k_1\over u}\Bigr]\Bigr)\Bigr) =
\delta_{S \eta} \, ,
\label{Qc-s1-1-eD}
\end{equation}
for the doublet states. To derive these $m\rightarrow 0$ and $n\rightarrow 1$
phase-shift expressions we used Eqs. (\ref{Phicc-cs1})-(\ref{Phicn-ssn1}) of 
Appendix A. For the phase shift of the $\eta$-spin singlet (and spin singlet) 
states we also used the two-pseudofermion phase-shift expression
given in Eq. (\ref{Phis-2}) of that Appendix. In the above overall phase-shift 
expressions (\ref{Qc-1-eT})-(\ref{Qc-s1-1-eD}), $k_1 = k^0 (q_1)$, $k_2 = k^0 (q_2)$, 
$\Lambda^{0}_{c0}(q)=\sin k^0 (q)$, ${\Lambda'}_1 = \Lambda^0_{s1} ({q'}_1)$, 
${\Lambda'}_2 = \Lambda^0_{s1} ({q'}_2)$, the rapidity functions $k^0 (q)$ and 
$\Lambda^0_{s1} (q)$ are the inverse of the functions defined by the first 
and second equations of Eq. (A.1) of Ref. \cite{V-1}, respectively, with 
$\nu =1$ in the second equation, the function $B(r)$ is defined in Eq. (\ref{Br}) 
of Appendix A, and $u=U/4t$. 

By inspection of the above phase-shift expressions one indeed confirms that 
$\pi + Q_{c0} (q_1)$ with $Q_{c0} (q_1)$ provided in Eqs. (\ref{Qc-1-eT}), (\ref{Qc-1-eS}),
and (\ref{Qc-s1-1-eD}) equals the phase shifts $\delta_{CT}$, $\delta_{CS}$, and
$\delta_{\eta S}$, respectively, given in Ref. \cite{S}. The two former phase shifts
are provided in Eq. (5.19) and the latter in Eq. (5.21) of that reference.
Moreover, the phase shift $Q_{s1} ({q'}_1)$ provided in Eqs. (\ref{Qs1-1-T}),
(\ref{Qs1-1-S}), and (\ref{Qc-s1-1-eD}) equals the phase shifts $\delta_{ST}$, 
$\delta_{SS}$, and $\delta_{S \eta}$, respectively, given in the same
reference. In this case the two former phase shifts are provided in Eq. (5.20) 
and the latter in Eq. (5.21) of Ref. \cite{S}. For the phase shifts of the doublet excited
states the confirmation of the above equalities also involves that $\arctan (\sinh(\pi x))=
2 \arctan (\exp (\pi x))-\pi/2$ for the branch such that these functions vary 
between $-\pi/2$ and $+\pi/2$.

\begin{table}
\begin{tabular}{|c||c|c|c|c|c|c|c|c|} \hline
Excited state & $\Delta N_{\uparrow}$ & $\Delta N_{\downarrow}$ &  
$\Delta M_{c,\,-1/2}$ & $\Delta M_{c,\,+1/2}$ & $\Delta M_{s,\,-1/2}$ & $\Delta M_{s,\,+1/2}$ & $Q^0_{c0}$ & $Q^0_{s1}$ \\
\hline\hline
$\eta$-spin triplet & -1 & -1 & 0 & 2 & -1 & -1 & $\pm\pi$ & $\pm\pi$ \\
\hline
$\eta$-spin triplet & 0 & 0 & 1 & 1 & -1 & -1 & $\pm\pi$ & $\pm\pi$ \\
\hline
$\eta$-spin triplet & 1 & 1 & 2 & 0 & -1 & -1 & $\pm\pi$ & $\pm\pi$ \\
\hline
$\eta$-spin singlet & 0 & 0 & 1 & 1 & -1 & -1 & 0 & $\pm\pi$ \\
\hline
spin triplet & 1 & -1 & 0 & 0 & -1 & 1 & $\pm\pi$ & $\pm\pi$ \\
\hline
spin triplet & 0 & 0 & 0 & 0 & 0 & 0 & $\pm\pi$ & $\pm\pi$ \\
\hline
spin triplet & -1 & 1 & 0 & 0 & 1 & -1 & $\pm\pi$ & $\pm\pi$ \\
\hline
spin singlet & 0 & 0 & 0 & 0 & 0 & 0 & $\pm\pi$ & 0 \\
\hline
doublet & 0 & 1 & 1 & 0 & 0 & -1 & $\pm\pi$ & 0 \\
\hline
doublet & -1 & 0 & 0 & 1 & 0 & -1 & $\pm\pi$ & 0 \\
\hline
doublet & 1 & 0 & 1 & 0 & -1 & 0 & $\pm\pi$ & 0 \\
\hline
doublet & 0 & -1 & 0 & 1 & -1 & 0 & $\pm\pi$ & 0 \\
\hline
\end{tabular}
\caption{The values of the spin-projection $\sigma$ electronic number
deviations $\Delta N_{\sigma}$, $\pm 1/2$ holon ($\alpha =c$) and 
$\pm 1/2$ spinon ($\alpha =s$) number deviations $\Delta M_{\alpha,\,\pm 1/2}$ 
for the pseudofermion representation of Refs. \cite{I,IIIb}, and scatter-less phase shifts  
$Q^0_{c0}$ and $Q^0_{s1}$ of  Eq. (\ref{pic0an}) of Appendix A for 
each class of excited states of Table I.}
\end{table}

Note that the phase shifts $\delta_{CT}$ and $\delta_{CS}$ given in
Eq. (5.19) and $\delta_{\eta S}$ in Eq. (5.21) of Ref. \cite{S} read
$\pi + Q_{c0} (q_1)$, whereas according to the phase-shift definition
of Eq. (\ref{qr1}) the corresponding $c0$ pseudofermion-hole phase
shifts are given by $Q_{c0} (q_1)$. The studies of Ref. \cite{S} used
the method of Ref. \cite{Korepin79} to evaluate the above phase
shifts.  That method provides the phase shifts up to an overall
constant term. In contrast, the method of Refs. \cite{S-P0,S-P}
provides the full corresponding pseudofermion-hole phase shift
value. In reference \cite{S} the term $\pi$ was added so that in the
limit $U\rightarrow\infty$ the phase factor $\exp \{i\delta_{CT}\}$
reads $\exp \{i\delta_{CT}\}=1$. However, the $c0$ pseudofermion and
hole phase shifts $Q_{c0} (q)$ of Eq. (\ref{qr1}) fully agree with the
corresponding $U\rightarrow\infty$ shifts used in the exact
one-electron spectral-function studies of Ref. \cite{Penc}. For the
scattering properties alone the constant extra term $\pi$ of the phase
shifts $\delta_{CT}$, $\delta_{CS}$, and $\delta_{\eta S}$ calculated
in Ref. \cite{S} is unimportant. In contrast, the use of the correct
overall pseudofermion-hole phase shift $Q_{c0} (q)$ defined as in
Eq. (\ref{qr1}) is required in the applications of the scattering
theory to the study of the finite-energy spectral properties
\cite{V-1,LE,Penc,spectral}.

A first important result for the clarification of the relation between the two 
representations is that in the reduced subspace the holon-scatterer and 
spinon-scatterer phase shifts of the conventional spinon-holon representation 
of Refs. \cite{Natan,S0,S} equal the $c0$ pseudofermion-hole and $s1$ 
pseudofermion-hole phase shifts of the pseudofermion representation
of Refs. \cite{S-P0,S-P}, respectively. 
(Except for $\pi$ for the holon scatterers.) Moreover, in this section we have 
shown that the phase shifts of the conventional spinon-holon representation 
are particular cases of the $c0$ and $s1$ pseudofermion overall phase-shift 
functionals $Q_{\alpha\nu}(q)$ defined by Eq. (\ref{Qcan1j}). The pseudofermion 
scattering theory refers to any initial ground state for densities in the ranges
$0\leq n\leq 1$ and $0\leq m\leq n$, whereas the spinon-holon scattering 
theory corresponds to the $n=1$ and $m=0$ initial ground state only. Furthermore, 
while the pseudofermion scattering theory is associated with a larger excitation 
subspace, which coincides with the PS, the phase shifts studied in Refs. 
\cite{Natan,S0,S} correspond to a reduced subspace spanned by the types of 
excited states of Tables I and II. 

\subsection{RELATION BETWEEN THE TWO CHOICES OF SCATTERING STATES AND 
CORRESPONDING $S$ MATRICES IN THE REDUCED SUBSPACE}

The above results enable us to relate the one-particle $S$ matrices of the two 
representations in the reduced subspace. Such an analysis provides useful 
information about the connection between the corresponding scattering states.
For the pseudofermion representation the expression of the unitary ${\hat{S}}$
operator which generates the virtual-state - "out"-state transition factorizes 
as ${\hat{S}} = \prod_{\alpha\nu}\prod_{j=1}^{N^*_{\alpha\nu}}
{\hat{S}}_{\alpha\nu ,q_j}$. Consistently, the excited energy eigenstates
can be written as a direct product of one-pseudofermion states.
It follows that application of the unitary ${\hat{S}}_{\alpha\nu ,q_j}$
operator onto a many-pseudofermion virtual state gives that state
multiplied by the $S$ matrix $S_{\alpha\nu} (q_j)$ given in Eq. (\ref{San}).
Thus, one can consider that both the many-pseudofermion virtual state 
and its one-pseudofermion state corresponding to the $\alpha\nu$ branch 
and bare-momentum $q_j$ are eigenstates of ${\hat{S}}_{\alpha\nu ,q_j}$
with the same eigenvalue $S_{\alpha\nu} (q_j)$. 

We have shown in the previous subsection that in the reduced subspace
spanned by the types of excited energy eigenstates of Tables I and II
the holon (and spinon) phase shift of the conventional spinon-holon 
representation equals except for $\pi$ (and equals) the corresponding
phase shift of the $c0$ pseudofermion-hole scatterer (and $s1$ 
pseudofermion-hole scatterer) corresponding to the same bare-momentum
value $q_j = 2\pi/I^{c0}_j$ (and $q_j = 2\pi/I^{s1}_j$). This reveals that
the $S$ matrix associated with the one-particle holon and spinon $S$ operator 
can also be obtained by application of the latter operator onto either the
one-particle state or the corresponding many-particle state.

Both the holons (and spinons) of the spinon-holon representation and 
the $c0$ pseudofermion holes (and $s1$ pseudofermion holes) correspond 
to the unoccupied quantum numbers $I^{c0}_j$ of the BA charge 
distribution of $k's$ excitations 
(of the unoccupied quantum numbers $I^{s1}_j$ of the BA spin string excitations of 
length one). However, that the holons (and spinons) of the spinon-holon 
representation and the $c0$ pseudofermion holes (and $s1$ pseudofermion 
holes) refer to the same unoccupied BA quantum numbers $I^{c0}_j$
(and $I^{s1}_j$) does not imply that their one-particle states are the same.
Indeed, the holon (and spinon) of momentum $q_j$ carries $\eta$-spin 
$1/2$ (and spin $1/2$), whereas the corresponding $c0$ pseudofermion hole
(and $s1$ pseudofermion hole) of momentum $q_j$ is a $\eta$-spin-less 
and spin-less object (and is a spin-zero object), as further discussed in
Sec. IV. Therefore, the relation of the one-pseudofermion scattering states 
to the holon or spinon one-particle scattering states of the conventional 
spinon-holon representation is a complex problem.

Fortunately, useful information about the relation between the one-particle
scattering states of both representations can be obtained by studying
the connection between the set of corresponding many-particle excited
states of the two representations with the 
precisely the same occupancy configurations of the BA
$I^{c0}_j$ and $I^{s1}_j$ quantum numbers. For each of the 
two alternative representations we replace the one-particle state under 
consideration by a suitable many-particle excited state with the same 
eigenvalue for the one-particle $S$ operator and thus with the same value for
the one-particle $S$ matrix. The relation between the many-particles states
of both representations is easier to achieve and provides important information
about the corresponding one-particle scattering states. 

Within the spinon-holon representation of Refs. \cite{S0,S}, the scatters and 
scattering centers are the $\pm 1/2$ spinons and $\pm 1/2$ holons created 
under the ground-state - excited state transitions. For the reduced subspace 
considered in these references, this leads to a spinon-spinon $4\times 4$ $S$ 
matrix, a holon-holon $4\times 4$ $S$ matrix, and a related $16\times 16$ $S$ 
matrix for the full scattering problem, as explained below. 
The holon-holon $4\times 4$ $S$ matrix (and spinon-spinon $4\times 4$ $S$ matrix) 
corresponds to the subspaces spanned by the four types of $\eta$-spin triplet and
singlet excited energy eigenstates (and spin triplet and singlet excited states) 
considered in Tables I and II. In this case the two objects 
created under the ground-state - excited-state
transitions are holons (and spinons) and thus it is unimportant 
which of them is chosen as scatterer and scattering center, since 
the phase shifts are the same. Thus, if one considers fixed momentum
values for the two created objects the number of relevant one-particle
scattering states equals that of excited energy eigenstates. In turn, for each
of the above considered four types of doublet excited energy eigenstates,
one must consider two one-particle scattering states. Indeed, as given in Table I, 
in this case one holon and one spinon are created under the 
ground-state - excited-state transitions and thus the one-particle
scattering states where that holon and spinon is the scatterer 
are different: the holon and spinon scattering states are associated
with different phase shifts which refer to the $c0$ and $s1$ phase 
shifts of Eq. (\ref{Qc-s1-1-eD}), respectively. 
Therefore, while at fixed momentum values of the two created objects
the reduced subspace is spanned by the twelve excited energy eigenstates
considered in Table I, the $S$ matrix for the corresponding full
scattering problem involves sixteen states and thus has dimension 
$16\times 16$. However, for each pair of one-particle
scattering states associated with the same doublet energy
eigenstate ones uses the latter state in the evaluation of
the corresponding phase shifts which appear in the entries 
of that $S$ matrix. It is of the form,
{\bear {\bf S} =\left[
\begin{array}{cccc}
{\bf S}_{SS} &0 &0 &0 \\
0 &{\bf S}_{S\eta} &0 &0 \\
0 &0 &{\bf S}_{\eta S} &0 \\
0 &0 &0 &{\bf S}_{CC} \label{MS}
\end{array}\right]\, ,
\ear}where ${\bf S}_{S\eta}$ and ${\bf S}_{\eta S}$ are $4\times 4$ diagonal matrices
corresponding to scattering events where the spinons and holons are the
scatterers and the holons and spinons the scattering centers, respectively.
In turn, ${\bf S}_{SS}$ and ${\bf S}_{CC}$ are the above two $4\times 4$ $S$ 
matrices associated with the spinon-spinon and holon-holon scattering, respectively.
The latter two matrices are not diagonal. We denote the corresponding two
sets of four states which correspond to the four one-particle scattering states 
by $\vert +1/2,+1/2;\alpha\rangle$,
$\vert -1/2,-1/2;\alpha\rangle$, $\vert +1/2,-1/2;\alpha\rangle$, and
$\vert -1/2,+1/2;\alpha\rangle$, where $\alpha = c$ and $\alpha = s$ refer
to the holon-holon and spinon-spinon states, respectively. The 
two $\eta$-spin ($\alpha = c$) or spin ($\alpha = s$) projections  $\pm 1/2$ 
labeling these states are those of the two involved holons or spinons,
respectively. The $4\times 4$ permutation matrix ${\bf P}$ transforms these 
four states as,
\begin{eqnarray}
\vert +1/2,+1/2;\alpha\rangle\hspace{0.50cm} & \Longrightarrow & \hspace{0.50cm}\vert +1/2,+1/2;\alpha\rangle \, ; \hspace{0.5cm}
\vert +1/2,-1/2;\alpha\rangle\hspace{0.50cm} \Longrightarrow \hspace{0.50cm}\vert -1/2,+1/2;\alpha\rangle \, , \nonumber \\
\vert -1/2,+1/2;\alpha\rangle\hspace{0.50cm} & \Longrightarrow & \hspace{0.50cm}\vert
+1/2,-1/2;\alpha\rangle \, ; \hspace{0.5cm}
\vert -1/2,-1/2;\alpha\rangle\hspace{0.50cm} \Longrightarrow \hspace{0.50cm}\vert -1/2,-1/2;\alpha\rangle
\, ; \hspace{0.25cm} \alpha = c,\,s \, . \label{scstates}
\end{eqnarray}
The two above non-diagonal matrices ${\bf S}_{SS}$ and ${\bf S}_{CC}$ are
then of the following form,
\begin{equation}
{\bf S}_{\beta\beta} = {1\over 2}(S_{\beta T}+S_{\beta S})\,{\bf I} +{1\over
2}(S_{\beta T}-S_{\beta S})\,{\bf P} \, ; \hspace{0.25cm} \beta = C,\,S \, ,
\label{MSbb}
\end{equation}
where ${\bf I}$ is the $4\times 4$ unity matrix, 
\begin{equation}
S_{\beta \tau} = (-1)^{y_{\beta}}\,e^{i\delta_{\beta\tau}} \, ; \hspace{0.5cm} \beta =C,
S \, ; \hspace{0.5cm} \tau =T, S \, ; \hspace{0.5cm} y_C = 1 \, , \hspace{0.25cm} y_S = 0
\, , \label{Sbt}
\end{equation}
and $\delta_{\beta\tau}$ with $\beta =C,\,S$ and $ \tau =T,\,S$ are the four phase
shifts defined by Eqs. (\ref{Qc-1-eT}), (\ref{Qc-1-eS}), (\ref{Qs1-1-T}), and 
(\ref{Qs1-1-S}). In turn, the four diagonal entries of the above $4\times 4$ diagonal matrices 
${\bf S}_{S\eta}$ and ${\bf S}_{\eta S}$ are equal and given by,
\begin{equation}
S_{\beta'\beta''} = (-1)^{z_{\beta'}}\,e^{i\delta_{\beta'\beta''}} \, ; \hspace{0.5cm}
\beta'\beta'' =S\eta ,\eta S \, ; \hspace{0.5cm} z_{\eta} = 1 \, , \hspace{0.25cm} z_S =
0 \, , \label{Sbb}
\end{equation}
where $\delta_{\beta'\beta''}$ with $\beta'\beta'' =S\eta,\,\eta S$ are the
two phase shifts defined in Eq. (\ref{Qc-s1-1-eD}). 

As discussed above, the pseudofermion-representation method for 
evaluation of phase shifts of Refs. \cite{S-P0,S-P} 
leads to the general phase-shift functional expression defined by Eqs. (\ref{qcan1j}) and
(\ref{Qcan1j}). Such a method provides the full phase-shift expressions. 
In contrast, the method of Ref. \cite{Korepin79} used in the studies of Refs. \cite{S0,S} 
provides the phase shifts (and corresponding $S$ matrices) of the
reduced-subspace excited states except for an overall constant term
(and an overall constant factor). This is behind a factor $-1$ appearing in the 
$S$ matrices given in Eqs. (\ref{Sbt}) and (\ref{Sbb}) for $\beta =C$ and $\beta' =\eta$, respectively,
relative to the corresponding $S$ matrices of Refs. \cite{S0,S}. 

Within the pseudofermion representation of Refs. \cite{S-P0,S-P}, we
denote the four excited energy eigenstates associated with the 
four one-pseudofermion scattering states corresponding to the four holon-holon
states $\vert +1/2,+1/2;c\rangle$, $\vert +1/2,-1/2;c\rangle$, $\vert -1/2,+1/2;c\rangle$, 
and $\vert -1/2,-1/2;c\rangle$ (and spinon-spinon states $\vert +1/2,+1/2;s\rangle$, 
$\vert +1/2,-1/2;s\rangle$, $\vert -1/2,+1/2;s\rangle$, and $\vert -1/2,-1/2;s\rangle$)
by $\vert c0,c0;-1\rangle$, $\vert c0,c0;0\rangle$, 
$\vert c0,c0,c1;0\rangle$, and $\vert c0,c0;+1\rangle$ (and
$\vert s1,s1;-1\rangle$, $\vert s1,s1;0\rangle$, 
$\vert s1,s1,s2;0\rangle$, and $\vert s1,s1;+1\rangle$.)
However, there is no one-to-one correspondence between the 
four states of each representation, as confirmed below. 
The index with values $0,\pm 1$ of these states refers 
to their value of $S_{c}^{z}$ (and $S_{s}^{z}$). The 
three states denoted by $\vert c0,c0;S_{c}^{z}\rangle$ (and
$\vert c0,c0;S_{c}^{z}\rangle$) correspond to $S_{c}^{z}=0,\pm 1$ 
(and $S_{s}^{z}=0,\pm 1$) and are the three $\eta$-spin triplet  
excited states (and three spin triplet  excited states) 
considered in Table I. Our analysis involves the phase shift
of the $c0$ (and $s1$) pseudofermion-hole scatterer. Indeed, the two branch 
indices $c0,c0$ (and $s1,s1$) of these states refer to the $c0$ (and $s1$)
pseudofermion-hole scatterer and $c0$ (and $s1$) pseudofermion-hole 
scattering center, respectively. In turn, 
$\vert c0,c0,c1;0\rangle$ (and $\vert s1,s1,s2;0\rangle$)
denotes the $\eta$-spin singlet excited state (and spin singlet excited state)
whose three branch indices $c0,c0,c1$ (and $s1,s1,s2$) refer to 
the $c0$ pseudofermion-hole scatterer, $c0$ pseudofermion-hole 
scattering center, and $c1$ pseudofermion scattering center
(and $s1$ pseudofermion-hole scatterer, $s1$ pseudofermion-hole 
scattering center, and $s2$ pseudofermion scattering center).

In contrast, the eight one-particle scattering states of Refs. \cite{S0,S}
associated with ground-state - excited-state transitions where 
one holon and one spinon are created correspond to
only the four doublet excited energy eigenstates 
considered above. Moreover, in that case the four
many-particle states of both representations associated
with these eight one-particle scattering states are the same states.
However, the absence of one-to-one correspondence between 
the other eight many-particle states associated with the
eight one-particle scattering states involving objects of the same
type (two spinons or two holons for the spinon-holon representation 
of Refs. \cite{S0,S}) and the corresponding
eight excited energy eigenstates implies that the $16\times 16$ $S$ 
matrix corresponding to the reduced-subspace scattering problem 
has a different form for the two representations. In the case of the pseudofermion
representation of Refs. \cite{S-P0,S-P}, we find for the reduced
subspace a diagonal $16\times 16$ $S$ matrix which is related 
to the non-diagonal $S$ matrix given in Eq. (\ref{MS}) by a unitary 
transformation as follows,
{\bear {\bf \bar{S}} = {\bf U}^{\dag}{\bf S}{\bf U} = \left[
\begin{array}{cccc}
{\bf \bar{S}}_{SS} &0 &0 &0 \\
0 &{\bf \bar{S}}_{S\eta} &0 &0 \\
0 &0 &{\bf \bar{S}}_{\eta S} &0 \\
0 &0 &0 &{\bf \bar{S}}_{CC} 
\end{array}\right] \, ; \hspace{0.5cm}
{\bf U} = \left[
\begin{array}{cccc}
{\bf J} &0 &0 &0 \\
0 &{\bf I} &0 &0 \\
0 &0 &{\bf I} &0 \\
0 &0 &0 &{\bf J} 
\end{array}\right]  \, ; \hspace{0.5cm}
{\bf J} =\left[
\begin{array}{cccc}
1 &0 &0 &0 \\
0 &1/\sqrt{2} &-1/\sqrt{2} &0\\
0 &1/\sqrt{2} &1/\sqrt{2} &0 \\
0 &0 &0 &1 \label{MS-U}
\end{array}\right]\, .
\ear}Here the matrix ${\bf U}$ is unitary. The three first diagonal entries (and
the fourth diagonal entry) of the two $4\times 4$ diagonal matrices 
${\bf \bar{S}}_{\beta\beta}$ such that $\beta = C,\,S$ of the above ${\bf \bar{S}}$ 
expression equal the phase factor $S_{\beta T}$ 
(and equals the phase factor $S_{\beta S}$) given in Eq. (\ref{Sbt}). 
The four diagonal entries of the other two $4\times 4$ diagonal matrices 
${\bf \bar{S}}_{\beta'\beta''}$ such that $\beta'\beta''
=S\eta ,\eta S$ of the same expression are equal and given in Eq. (\ref{Sbb}).
For the general pseudofermion scattering theory the 
diagonal entries of the $16\times 16$ diagonal matrix ${\bf \bar{S}}$ 
provided in Eq. (\ref{MS-U}) are the sixteen $S$ matrices 
$S_{\alpha\nu} (q_j)$ of dimension one and of general form given in Eq. (\ref{San}) 
corresponding to the $c0$ and $s1$ pseudofermion-hole scatterers of the 
reduced-subspace excited states considered here. 

Use of the unitary matrix defined in Eq. (\ref{MS-U}) 
reveals that the above four holon-holon ($\alpha =c$) and four spinon-spinon ($\alpha =s$)
states can be expressed in terms of the excited energy eigenstates which contain
the one-pseudofermion scattering states of the alternative representation as follows, 
\begin{eqnarray}
\vert +1/2,+1/2;\alpha\rangle & = & \vert \alpha\nu ,\alpha\nu;-1\rangle \, ; \hspace{0.5cm}
\vert +1/2,-1/2;\alpha\rangle =
{1\over\sqrt{2}}\Bigl[\vert \alpha\nu ,\alpha\nu;0\rangle -
\vert \alpha\nu ,\alpha\nu ,\alpha\nu';0\rangle\Bigr] \, , \nonumber \\
\vert -1/2,+1/2;\alpha\rangle & = & {1\over\sqrt{2}}\Bigl[\vert \alpha\nu ,\alpha\nu;0\rangle + 
\vert  \alpha\nu ,\alpha\nu ,\alpha\nu';0\rangle\Bigr] \, ; \hspace{0.5cm}
\vert -1/2,-1/2;\alpha\rangle = \vert \alpha\nu ,\alpha\nu;+1\rangle \, .
\label{scstates-U}
\end{eqnarray}
Here $\alpha\nu=c0,\,s1$ or $\alpha\nu'=c1,\,s2$, respectively.
This confirms that the states $\vert +1/2,-1/2;\alpha\rangle$ and 
$\vert -1/2,+1/2;\alpha\rangle$ associated with the holon-holon
and spinon-spinon one-particle scattering states of the conventional 
spinon-holon representation are not eigenstates of the $\eta$ 
spin ($\alpha =c$) or spin ($\alpha =s$). 

%%%%%%%%%%%%%%%%%%%%%%%%%%%%%%%%%%%%%%%%%%%%%%%%%%%%%%%%%%%%%%%%
\section{FAITHFUL CHARACTER AND SUITABILITY TO THE STUDY OF THE SPECTRAL 
PROPERTIES OF BOTH THEORIES IN THE PS}

In this section we consider the extension of the scattering theory associated
with the spinon-holon representation of Refs \cite{Natan,S0,S} 
to the whole PS and show that similarly to the pseudofermion scattering theory
it is faithful there. Furthermore, we discuss the suitability of 
the two representations under consideration for applications to the study of the 
finite-energy spectral and dynamical properties.

\subsection{FAITHFUL CHARACTER OF BOTH REPRESENTATIONS IN THE PS AND THE CHARGE
AND SPIN CARRIED BY THE CORRESPONDING QUANTUM OBJECTS}

The rotated-electron holon and spinon description introduced in Ref. \cite{I} was shown in that reference
to be a faithful representation for the whole Hilbert space. For the PS that the pseudofermion
description refers to, by faithful representation
we mean that for each subspace with fixed values $S_c$ of $\eta$ spin, $S$ of spin, $M_c$
of the holon number, and $M_s$ of the spinon number, the corresponding number of $\eta$-spin 
(and spin) irreducible representations equals the number of $\nu\geq 1$ composite
$c\nu$ pseudofermions and $-1/2$ Yang holons (and $\nu\geq 1$ composite $s\nu$
pseudofermions and $-1/2$ HL spinons) occupancy configurations of the energy eigenstates
that span such subspaces. The dimension of any PS subspace spanned by all
energy eigenstates with fixed values of $S_c$, $S_s$, $M_c$, and $M_s$ is given by \cite{I},
\begin{equation}
{N_a\choose N_{c0}}\times{\cal{N}}(S_c ,M_c)\times{\cal{N}}(S_s ,M_s)  \, . \label{LCS}
\end{equation}
Here ${\cal{N}}(S_c ,M_c)$ and ${\cal{N}}(S_s ,M_s)$ is the number of states with fixed
$\eta$-spin value $S_c$ and spin value $S_s$, respectively, representative of a
collection of a number $M_c$ of $\eta$-spin $1/2$ holons and $M_s$ of spin $1/2$ spinons,
respectively. The faithful character of this representation follows from the equality
of the following two numbers: The number given in Eq. (47) of Ref. \cite{I} of 
$\eta$-spin ($\alpha =c$) and spin ($\alpha =s$) irreducible representation states of $M_c$
$\eta$-spin $1/2$ holons and $M_s$ spin $1/2$ spinons, arranged within all possible
configurations with fixed $\eta$-spin value $S_c$ and spin value $S_s$, respectively,
and the number provided in (51) of that reference. The latter is the product of the number
discrete bare-momentum $\alpha\nu$ pseudofermion occupancy configurations such 
that the number of $2\nu$-holon 
composite $c\nu$ pseudfermions $(\alpha =c)$ or $2\nu$-spinon composite $s\nu$ 
pseudofermions $(\alpha =s)$ obey the sum rule $\sum_{\nu =1}^{\infty}\,\nu\,N_{\alpha\nu} = 
[M_{\alpha}/2-S_{\alpha}]$ by the number of possible occupancies of the $-1/2$ and $+1/2$ Yang
holons $(\alpha =c)$ and $-1/2$ and $+1/2$ HL spinons $(\alpha =s)$ such that
$L_{\alpha}=[L_{\alpha,\,+1/2}+ L_{\alpha,\,-1/2}]=2S_{\alpha}$. In reference 
\cite{I} it is shown that this equality occurs for all the above subspaces with fixed
values for $S_c$, $S_s$, $M_c$, and $M_s$. For the rotated-electron holon and spinon 
description of that reference the $\eta$-spin $SU(2)$ irreducible representations 
correspond to the BA charge string excitations of length $\nu =1,2,3,...$ and $\pm 1/2$ Yang holon
occupancy configurations. Moreover, the number ${N_a\choose N_{c0}}$ of $c0$
pseudofermion and hole occupancy configurations appearing in Eq. (\ref{LCS}) does not
count $\eta$-spin $SU(2)$ irreducible representations. This implies that within the 
rotated-electron holon and spinon definition of 
Ref. \cite{I} the occupancy configurations of the $c0$
pseudofermion and holes of states belonging to the PS are independent of the $\eta$-spin
degrees of freedom and thus are not related to the $\eta$-spin $1/2$ holons.
Moreover, note that according to Eq. (51) of Ref. \cite{I} with $\alpha =s$ the 
number of occupancy configurations ${N_{s1}+N_{s1}^h\choose
N_{s1}}$ of the $s1$ pseudofermions and holes (holes of the length-one spin string
excitation spectrum) contribute to the number of spin singlet representation states with
fixed $S_{s}\leq M_{s}/2$ value which according to the
spin summation rules one can generate from $M_{s}$ spin $1/2$ spinons. This  
is consistent with the spin singlet character of the
$N_{s1}$ two-spinon composite $s1$ pseudofermions and $N_{s1}^h$ $s1$ pseudofermion
holes. 

Let us next consider the spinon and holon definition of the conventional
spinon-holon representation of Refs. \cite{Natan,S0,S}. For the reduced 
subspace considered in the previous section, 
the spin $1/2$ spinons are identified with
the holes of the length-one BA spin string excitation spectrum. Moreover, the holes of the BA
distribution of $k's$ excitation spectrum 
are identified with single $\eta$-spin $1/2$ holons.
In order to confirm the faithful character of the spinon-holon representation of Refs.
\cite{Natan,S0,S} and search whether it is suitable for the description of the
finite-energy spectral and dynamical properties of the metallic phase, 
it is convenient to extend it to the whole PS and to initial 
ground states corresponding to densities in the ranges $0<n<1$ and $0<m<n$. 
In the remaining of this paper we call it {\it extended} spinon-hole representation
or theory.

The extended spinon-hole representation assumes that the $N^h_{c0}$ holes in the BA distribution of
$k's$ excitation spectrum are $\eta$-spin $1/2$ holons and the $N^h_{s1}$ holes in the
length-one BA spin string excitation spectrum are spin $1/2$ spinons. Thus, for
electronic densities $n<1$ and spin densities $m>0$ the initial ground state itself has a
finite number of holons and spinons. Each ground-state - excited-state transition leads
to new values $N^h_{c0}+\Delta N^h_{c0}$ and $N^h_{s1} + \Delta N^h_{s1}$. For the PS the
deviations $\Delta N^h_{c0}$ and $\Delta N^h_{s1}$ refer to a finite number of created
holons and spinons, respectively. The excited states considered in Sec. III 
and in Refs. \cite{S0,S} correspond to a particular case of this extended spinon-holon
theory where $N^h_{c0}=N^h_{s1}=0$ for the initial ground state and $\Delta N^h_{c0}+
\Delta N^h_{s1}=2$ for the excited states. Let us denote the number of holons and 
spinons of such an extended theory by ${\cal M}_c\equiv N^h_{c0}$ and 
${\cal M}_s\equiv N^h_{s1}$, respectively.
The relation between the numbers of holons and spinons of both representations is such that, 
\begin{equation}
{\cal M}_c={\cal M}_{c,\,+1/2} + {\cal M}_{c,\,-1/2} = M_c \, ; \hspace{0.5cm} {\cal M}_s= 
{\cal M}_{s,\,+1/2} + {\cal M}_{s,\,-1/2} =M_s - 2\sum_{\nu
=1}^{\infty}N_{s\nu} \, , \label{Nhs1}
\end{equation}
and thus ${\cal M}_s< M_s$. Here ${\cal M}_{\alpha,\,\pm 1/2}$  denotes the number 
of $\eta$-spin-projection $\pm 1/2$ 
holons ($\alpha =c$) and spin-projection $\pm 1/2$ spinons ($\alpha =s$). As for the
other representation, we call these objects $\pm 1/2$ holons and $\pm 1/2$ spinons,
respectively. Since the spinon-holon representation 
of Refs. \cite{Natan,S0,S} corresponds to one-particle scattering 
states which are part of eigenstates of the diagonal generators of the $\eta$-spin and spin algebras, 
the above numbers ${\cal M}_{\alpha,\,\pm 1/2}$ are related to the eigenvalues of these 
generators as follows,
\begin{equation}
-2 S_{\alpha}^z = {\cal M}_{\alpha,\,+1/2} - {\cal M}_{\alpha,\,-1/2} \, ; \hspace{0.5cm}
\alpha = c,\, s\, . \label{Ma}
\end{equation}
Within the extended spinon-holon scattering theory the  $\pm 1/2$ spinons and $\pm 1/2$ holons
are the scatters and scattering centers. Thus, it follows from the finite spin and $\eta$-spin value
of such scatterers and scattering centers that some of the corresponding one-particle scattering states 
do not correspond to eigenstates of the total spin and $\eta$-spin, 
as confirmed in the previous section for states belonging 
to the reduced subspace. Moreover, we find below that, in contrast to the corresponding one-pseudofermion
scattering states, for initial ground states with densities $n<1$ and/or $m>0$ some of the 
one-particle scattering states of the extended spinon-holon theory 
do not correspond to energy and momentum eigenstates. 

By combining Eq. (\ref{Ma}) with the relations given in Eq. (\ref{Nhs1})
we find that,
\begin{equation}
{\cal M}_{c,\,\pm 1/2} = M_{c,\,\pm 1/2} \, ; \hspace{0.5cm} {\cal M}_{s,\,\pm 1/2} =
M_{s,\,\pm 1/2} - \sum_{\nu =1}^{\infty}N_{s\nu} \, . \label{M-M}
\end{equation}
However, the holon number equality ${\cal M}_{c,\,\pm 1/2} = M_{c,\,\pm 1/2}$ does not 
imply that the $\pm 1/2$ holons of the extended spinon-holon representation are 
the same quantum objects as those of the pseudofermion representation, 
as confirmed below. Indeed, the holons of 
both representations have different expressions in terms of rotated electrons and 
thus transform differently under the electron - rotated-electron unitary transformation 
and carry a different elementary charge. Furthermore, in contrast to the holons of the 
pseudofermion representation, those of the extended spinon-holon representation 
have a momentum-dependent energy dispersion.

Let us confirm that the extended spinon-holon representation is
also faithful in the PS.  The dimension of a PS subspace spanned by all energy
eigenstates with fixed values of $S_c$, $S_s$, ${\cal M}_c$, and ${\cal M}_s$
as given in Eq. (\ref{Nhs1}) reads,
\begin{equation}
{N_a\choose {\cal M}_c}\times{N^*_{s1}\choose {\cal M}_s}\times{\cal{N}}(S_c ,{\cal M}_c)\times
{\cal{N}}(S_s ,{\cal M}_s)  \, . \label{LCS*}
\end{equation}
Here ${\cal{N}}(S_c ,{\cal M}_c)$ and ${\cal{N}}(S_s ,{\cal M}_s)$ is the number of states
with fixed $\eta$-spin value $S_c$ and spin value $S_s$, respectively, representative of
a collection of a number ${\cal M}_c$ of $\eta$-spin $1/2$ holons and ${\cal M}_s$ of spin
$1/2$ spinons, respectively. We emphasize that the PS subspaces with fixed values for $S_c$, $S_s$,
${\cal M}_c$, and ${\cal M}_s$ are smaller than those with fixed values for $S_c$, $S_s$,
$M_c$, and $M_s$, as confirmed below. The faithful character of the alternative extended
spinon-holon representation requires that the numbers $S_c$, $S_s$,
${\cal M}_c$, and ${\cal M}_s$ must obey the following equality,
\begin{equation}
{\cal{N}}(S_{\alpha} ,{\cal M}_{\alpha}) = (2S_{\alpha} +1)\left\{
{{\cal M}_{\alpha}\choose {\cal M}_{\alpha}/2-S_{\alpha}} - {{\cal M}_{\alpha}\choose
{\cal M}_{\alpha}/2-S_{\alpha}-1}\right\} = (2S_{\alpha} +1)\, \sum_{\{N_{\alpha\nu'}\}}\,
\prod_{\nu' =1+x_{\alpha}}^{\infty}\,{N_{\alpha\nu'}+N_{\alpha\nu'}^h\choose N_{\alpha\nu'}} 
\, , \label{Nst1*}
\end{equation}
where $ \alpha = c ,s$, the sum and product $\sum_{\{N_{\alpha\nu'}\}}\,\prod_{\nu' =1}^{\infty}$
run over occupancies such that the number $\sum_{\nu'=1+x_{\alpha}}^{\infty}\,(\nu'-x_{\alpha})\,N_{\alpha\nu'} =
[{\cal M}_{\alpha}/2-S_{\alpha}]$ is fix, $x_c = 0$, and $x_s = 1$. 
The factor ${{\cal M}_{\alpha}\choose {\cal M}_{\alpha}/2-S_{\alpha}} - {{\cal M}_{\alpha}\choose
{\cal M}_{\alpha}/2-S_{\alpha}-1}$ in this equation is the number of $\eta$-spin ($\alpha =c$) 
and spin ($\alpha =s$) singlet representation
states with fixed $S_{\alpha}\leq {\cal M}_{\alpha}/2$ value which according to the
$\eta$-spin and spin summation rules one can generate from ${\cal M}_{\alpha}$ quantum
objects of $\eta$ spin $1/2$ and spin $1/2$, respectively. By multiplying this number by
the number $(2S_{\alpha} +1)$ of states in each $SU(2)$ tower, one reaches the number of
$\eta$-spin ($\alpha =c$) and spin ($\alpha =s$) irreducible representation states of
${\cal M}_c$ $\eta$-spin $1/2$ holons and ${\cal M}_s$ spin $1/2$ spinons, arranged within
all possible configurations with fixed $\eta$-spin value $S_c$ and spin value $S_s$,
respectively. In turn, the quantity $\sum_{\{N_{\alpha\nu'}\}}\, \prod_{\nu'
=1+x_{\alpha}}^{\infty}\,{N_{\alpha\nu'}+N_{\alpha\nu'}^h\choose N_{\alpha\nu'}}$
on the right-hand side of Eq. (\ref{Nst1*}) gives the number of occupancy configurations 
of the BA quantum numbers such that the sum rule $\sum_{\nu' =1+
x_{\alpha}}^{\infty}\,(\nu'-x_{\alpha})\,N_{\alpha\nu'} =[{\cal M}_{\alpha}/2-S_{\alpha}]$ 
is obeyed. Furthermore, $(2S_{\alpha} +1)$ refers to the tower of states outside the 
BA solution. The point is that the equality (\ref{Nst1*}) is indeed valid for all PS 
subspaces with fixed values for $S_c$, $S_s$, ${\cal M}_c$, and ${\cal M}_s$. 
Thus, the spinon-holon representation is faithful both for the PS and the reduced 
subspace spanned by the types of excited states of Tables I of Sec. III.

For each of the subspaces with fixed values of $S_c$, $S_s$, $M_c$, and $M_s$
associated with the $c0$ pseudofermion, holon, and spinon 
representation of Refs. \cite{I,IIIb}, the subspace
dimension of Eq. (\ref{LCS}) is a product of three numbers. Two of these numbers are
nothing but the value given in Eq. (47) of Ref. \cite{I} of different states with the same value of
$S_{\alpha}$ that, following the counting rules of $\eta$-spin and spin summation, one
can generate from $M_{\alpha}$ $\eta$-spin $1/2$ holons ($\alpha =c$) and spin $1/2$
spinons ($\alpha =s$). These two values are uniquely defined by the fixed values of the
total $\eta$ spin and spin of the subspace and by the fixed numbers of $\eta$-spin $1/2$
holons and spin $1/2$ spinons in that subspace. The third number corresponds to the $c0$
pseudofermion excitations. This is the number of states associated with the possible
occupancy configurations of $N_{c0}$ $c0$ pseudofermions and $N^h_{c0}$ $c0$
pseudofermion holes where $N_a=N_{c0}+N^h_{c0}$. These charge excitations describe the
translational motion of the rotated-electron singly-occupied sites relative to the
rotated-electron doubly-occupied and empty sites.

In turn, the subspace dimension of Eq. (\ref{LCS*}) associated with the extended
spinon-holon representation is a product of four factors. 
Each subspace with fixed values of $S_c$, $S_s$, $M_c$, and $M_s$ contains several
subspaces with fixed values for $S_c$, $S_s$, ${\cal M}_c$, and ${\cal M}_s$.
Thus, the dimension (\ref{LCS*}) of a given PS subspace with fixed values of
$S_c$, $S_s$, ${\cal M}_c$, and ${\cal M}_s$ is smaller than the dimension (\ref{LCS}) 
of the larger subspace where it is contained. Two 
of the four factors of the dimension  (\ref{LCS*}) are
the value of Eq. (\ref{Nst1*}) of different states with the same value of $S_{\alpha}$
that, following the counting rules of $\eta$-spin and spin summation, one can generate
from ${\cal M}_{\alpha}$ $\eta$-spin $1/2$ holons ($\alpha =c$) and spin $1/2$ spinons
($\alpha =s$). These two values are uniquely defined by the fixed values of the total
$\eta$ spin and spin of the subspace and by the fixed numbers of $\eta$-spin $1/2$ holons
and spin $1/2$ spinons in that subspace. The other two factors refer to the
different choices of momentum occupancy configurations of the ${\cal M}_c$ holons and
${\cal M}_s$ spinons, respectively. (Such configurations correspond to the 
BA distribution of $k's$ for the holons and BA spin string excitations of length
one for the spinons.)  Indeed, in the extended spinon-holon representation the 
spinons and holons have momentum-dependent energy dispersions.

The faithful character of the extended spinon-holon representation
is closely related to the faithful character of the pseudofermion 
representation. As a matter of fact, the number given in Eq. (47) of Ref. \cite{I} 
of spin singlet representation states with fixed $S_{s}$ and $M_{s}$ 
values which according to the spin summation rules one can
generate from $M_{s}$ spin $1/2$ spinons of that reference can be expressed as the
following summation over the numbers of spin singlet representation
states with fixed $S_{s}$ value but different numbers ${\cal M}_s$ of spin $1/2$ spinons of
the alternative extended spinon-holon representation,
\begin{equation}
\left\{ {M_{s}\choose M_{s}/2-S_{s}} - {M_{s}\choose M_{s}/2-S_{s}-1}\right\} =
\sum_{\{{\cal M}_s\}} {N^*_{s1}\choose {\cal M}_s}\left\{ {{\cal M}_s\choose
{\cal M}_s/2-S_{s}} - {{\cal M}_s\choose {\cal M}_s/2-S_{s}-1}\right\}\, .
\label{relation-singlet}
\end{equation}
Here $M_s$ and $S_s$ are fixed and for each allowed value of ${\cal M}_s$ there is one and
only one value of $N_{s1}$ such that ${\cal M}_s=2S_s + 2\sum_{\nu'
=2}^{\infty}\,(\nu'-1)\,N_{s\nu'}$ and $N_{s1}=[M_s/2-S_s] - \sum_{\nu'
=2}^{\infty}\,\nu'\,N_{s\nu'}$, respectively. Thus, the summation on the right-hand side
of Eq. (\ref{relation-singlet}) is over the dimensions of all subspaces
with fixed values for $S_c$, $S_s$, ${\cal M}_c$, and ${\cal M}_s$ that are contained in a
single subspace with fixed values of $S_c$, $S_s$, $M_c$, and $M_s$.

An interesting point is the following. For the extended spinon-holon
representation the occupancy configurations whose 
number reads ${N^*_{s1}\choose {\cal M}_s} = {N^*_{s1}\choose N_{s1}}$
do not correspond to irreducible representations of the spin $SU(2)$ algebra, but instead
refer to the momentum occupancy configurations of the ${\cal M}_s$ spinons over the
available $N^*_{s1}$ discrete spin-rapidity momentum values, as confirmed by Eqs.
(\ref{LCS*}) and (\ref{Nst1*}). In contrast, for the pseudofermion representation 
the factor ${N^*_{s1}\choose {\cal M}_s} $ contributes to the number of irreducible
representations of the spin $SU(2)$ algebra associated with the $M_s$ spinons of spin
$1/2$. Moreover, while for the former representation ${N_a\choose  {\cal M}_c}={N_a\choose
N_{c0}}$ gives the number of momentum occupancy configurations of the ${\cal M}_c$ holons
over the available $N_a$ discrete distribution of $k'$ values, for the latter description
${N_a\choose N^h_{c0}}={N_a\choose N_{c0}}$ is the number of momentum 
occupancy configurations of the $N_{c0}$ $c0$ pseudofermions over the available 
$N_a$ discrete bare-momentum values. In contrast to the holons of 
the extended spinon-holon representation, the $c0$ pseudofermions and holes
have no $\eta$-spin degrees of freedom \cite{I,IIIb}. 
On the other hand, for the $\alpha\nu\neq c0,\,s1$ excitations the number of
occupancy configurations ${N^*_{\alpha\nu}\choose N_{\alpha\nu}}$ contributes to the numbers
of irreducible representations of the $\eta$-spin ($\alpha\nu =c\nu\neq c0$) and spin
($\alpha\nu =s\nu\neq s1$) $SU(2)$ algebras associated with the holons of $\eta$ spin
$1/2$ and spinons of spin $1/2$, respectively, of both representations.

Next, let us consider the transport of charge and spin within
the two alternative representations. The electronic charge and 
spin remain invariant under the 
electron - rotated-electron unitary transformation and thus the 
rotated electrons have the same charge and spin  as
the electrons \cite{I}. Thus, it follows from the relation
of the rotated electrons to the $-1/2$ holons and $+1/2$ 
holons that the latter objects carry charge $-2e$ 
and $+2e$, respectively. However, within the pseudofermion
representation only the $-1/2$ holons of 
charge $-2e$ are active charge carriers for the description 
of the charge transport in terms of electrons. In turn, for 
the description of the charge transport in terms of electronic 
holes, only the $+1/2$ holons of charge $+2e$ are active 
charge carriers. The charge is also carried by the $c0$ 
pseudofermions, which describe the charge degrees of 
freedom of the lattice sites singly occupied by rotated electrons. 
As discussed in Refs. \cite{I,IIIb}, for the description of the charge 
transport in terms of electrons (and electronic holes) the 
$c0$ pseudofermions carry charge $-e$  (and $+e$). (Such
objects have no spin and $\eta$-spin degrees of freedom.)
We recall that the $c\nu\neq c0$ pseudofermions (and $s\nu$ 
pseudofermions) are $\eta$-spin zero (and spin zero) composite objects 
of an equal number $\nu=1,2,...$ of $-1/2$ holons and $+1/2$ holons 
(and $-1/2$ spinons and $+1/2$ spinons). Thus, within the description of 
charge transport in terms of electrons (and electronic holes), the $c\nu$ 
pseudofermions carry charge $-2\nu e$ (and $+2\nu e$) where $\nu =1,2,...$. 

The charge $-e$ carried by the $\Delta N$ electrons (or the charge $+e$ 
carried by the $\Delta N^h$ electronic holes) involved in a transition 
from the ground state to an excited state is distributed by the objects 
of the pseudofermion respresentation as given in Eq. (57) of Ref. \cite{I}. 
Also the electronic spin remains invariant under the electron - rotated-electron 
unitary transformation. Thus, within the pseudofermion representation, 
the deviations $\Delta N_{\uparrow}$ and $\Delta N_{\downarrow}$ 
in the numbers of electronic up 
and down spins, respectively, are distributed by the quantum objects as 
$\Delta N_{\uparrow} = \Delta M_{s,\,+1/2} + \Delta M_{c,\,-1/2} = 
\sum_{\nu=1}^{\infty} \nu\,\Delta N_{s\nu} +
\Delta L_{s,\,+1/2} + \sum_{\nu=1}^{\infty} \nu\,\Delta N_{c\nu} + \Delta L_{c,\,-1/2}$
and $\Delta N_{\downarrow} = \Delta M_{s,\,-1/2} + \Delta M_{c,\,-1/2} = 
\sum_{\nu=1}^{\infty} \nu\,\Delta N_{s\nu} +
\Delta L_{s,\,-1/2} + \sum_{\nu=1}^{\infty} \nu\,\Delta N_{c\nu} + \Delta L_{c,\,-1/2}$.
Note that in addition to the spinons, which correspond to the electronic
spins of the rotated-electron singly occupied sites, some of the electronic
spins refer to the rotated-electron doubly occupied sites. The latter
electronic spins are contained in the $-1/2$ holons. Each of these
objects corresponds to one spin-zero on-site pair of rotated electrons with opposite 
spin projection. Thus, in spite of the $-1/2$ holon being a spin-zero
object, it contains one electronic up spin and one electronic down spin. 
It follows that one $c\nu\neq c0$ pseudofermion, which is a composite
object of $\nu$ $-1/2$ holons and $\nu$ $+1/2$ holons, contains $\nu$ 
electronic up spins and $\nu$ electronic down spins. 
In spite of the Yang holons having charge and the HL spinons spin, such
objects have a localized character and thus do not contribute to the
transport of charge and spin, respectively \cite{I}.

For the extended spinon-holon theory the holons and spinons are behind
the transport of charge and spin, respectively.  The holons of that theory
carry half of the charge of those of the pseudofermion representation. 
Thus, such $-1/2$ holons and $+1/2$ 
holons carry charge $-e$ and $+e$, respectively \cite{S0,S}. Moreover, while 
for the pseudofermion representation the $-1/2$ holons and $+1/2$ holons 
correspond to alternative descriptions of the charge transport in terms of electrons 
and electronic holes, respectively, for the extended spinon-holon theory the charge
transport is performed at the same time by the $-1/2$ holons and $+1/2$ holons. 
Indeed, it follows from Eq. (\ref{Ma}) that for the latter theory one has that 
$(-e)\,[N-N_a] = (+e)\,[N^h-N_a] = (-e)\,{\cal M}_{c,\,-1/2}+(+e)\,{\cal M}_{c,\,+1/2}$ 
and thus the corresponding general ground-state - excited-state deviations
associated with the transport of charge are such that,
\begin{equation}
(-e)\,\Delta N= (-e)\,\Delta {\cal M}_{c,\,-1/2}+(+e)\,\Delta {\cal M}_{c,\,+1/2} \, .
\label{-eDN}
\end{equation}
Furthermore, concerning the spin  transport it also follows from Eq. (\ref{Ma}) that 
$N_{\uparrow} -N_{\downarrow}= {\cal M}_{s,\,+1/2} -{\cal M}_{s,\,-1/2}$ and thus
$\Delta N_{\uparrow} -\Delta N_{\downarrow}= \Delta {\cal M}_{s,\,+1/2} -\Delta {\cal
M}_{s,\,-1/2}$.

In conclusion, the two alternative representations associated with the scattering
theories of Refs. \cite{S-P0,S-P} and \cite{Natan,S0,S}, respectively, are faithful.
However, such scattering theories refer to two alternative choices of 
one-particle scattering states, scatterers, scattering centers, and carriers 
of charge and spin.

\subsection{THE EXTENDED SPINON-HOLON THEORY AND 
SUITABILITY FOR THE DESCRIPTION OF THE 
FINITE-ENERGY SPECTRAL AND DYNAMICAL PROPERTIES}

For the extended spinon-holon scattering theory 
the scatterers and (scattering centers) are 
the holons and spinons of the excited energy eigenstates 
associated with the one-particle scattering states 
(and the holons and spinons created 
under the corresponding ground-state - excited-state transitions).  
(Since the initial ground state of the reduced 
subspace considered in Sec. III and Refs.  \cite{S0,S} has no
holons and no spinons, all holons and spinons of the corresponding
excited states are both scatterers and scattering centers.)
Furthermore, for the extended spinon-holon 
representation the number of holons and
spinons equals the number of $c0$ pseudofermion holes and $s1$
pseudofermion holes, respectively, of the pseudofermion representation.
As for the reduced-subspace holon and spinon phase shifts 
studied in Sec. III, the holon-scatterer and spinon-scatterer phase shifts of the 
extended spinon-holon theory 
equal the corresponding phase shifts of the $c0$ pseudofermion holes and
$s1$ pseudofermion holes, respectively. The many-particle
states associated with the one-particle scattering states of that extended theory can 
always be expressed in terms of the excited energy eigenstates
associated with the corresponding one-pseudofermion scattering states, 
as we have illustrated in Sec. III for the reduced subspace. 
As for that reduced subspace, many of the one-particle scattering states of 
the extended spinon-holon theory do not correspond to
$\eta$-spin and spin eigenstates. On the other hand, the scattering states 
of such an extended scattering theory always refer to eigenstates of the 
$\eta$-spin and spin projections, but for initial ground states with densities 
in the ranges $0<n<1$ and $0<m<n$ many of these states 
do not correspond to energy and momentum eigenstates. The many-particle
states and associated one-particle scattering states 
of the two representations have the following general properties:
\begin{enumerate}

\item All one-pseudofermion scattering states correspond to 
excited energy and momentum eigenstates;

\item All excited many-particle states of one-particle scattering states of 
the extended spinon-holon theory whose expressions in terms 
of the excited energy eigenstates do not involve states with finite 
occupancy of $\alpha\nu$ pseudofermions belonging to 
$\alpha\nu\neq c0,\,s1$ branches are energy and momentum 
eigenstates;

\item For initial ground states with electronic density $n=1$ (and spin density
$m=0$) all excited many-particle states of one-particle scattering states of 
the extended spinon-holon theory whose expressions in terms 
of the excited energy eigenstates involve states with finite 
occupancy of $\alpha\nu$ pseudofermions belonging to
$c\nu\neq c0$ branches (and $s\nu\neq s1$ branches) are energy 
eigenstates;

\item For initial ground states with electronic densities in the
range $0<n<1$ (and spin densities in the range
$0<m<n$) the excited many-particle states of one-particle scattering states of 
the extended spinon-holon theory whose expressions in terms 
of the excited energy eigenstates involve states with finite 
occupancy of $\alpha\nu$ pseudofermions belonging to
$c\nu\neq c0$ branches (and $s\nu\neq s1$ branches) are not 
in general energy and momentum eigenstates;
\end{enumerate}

These properties refer to the Hamiltonian $\hat{H}$ of Eq. (\ref{H}),
whose excited-state energy is measured relative to that of the
initial ground state, as for the PDT spectral-function expressions
\cite{V-1,LE}. (Property 3 is valid when for initial ground states with 
electronic density $n=1$ the zero-energy level corresponds
to the middle of the Mott-Hubbard gap.) As a simple example, let us 
consider that the initial ground state has an electronic density in the 
range $0<n<1$ and such that $N$ is even and $N_{\uparrow}$
and $N_{\downarrow}$ are odd. In contrast to the $n=1$ ground
state, the $c0$ band of such a state is occupied by $c0$ 
pseudofermion holes for bare-momentum values in the range 
$2k_F<\vert q\vert <\pi$. Let us consider four excited 
energy eigenstates whose $c0$ pseudofermion occupancy configuration
differs from that of the initial ground state by the creation of two
$c0$ pseudofermion holes at given fixed bare momentum values $q_1$
and $q_2$ in the range $q_1,\,q_2\in [-2k_F,\,+2k_F]$ and such
that $q_1\neq q_2$. These four excited states are a generalization
for $n<1$ of the set of four excited states of the $n=1$ ground
state including the three $\eta$-spin triplet excited states 
and the $\eta$-spin singlet excited state considered  in Sec. III. As in 
that section, for each of the two alternative representations we replace 
the one-particle state under consideration by a suitable many-particle 
excited state with the same eigenvalue for the one-particle $S$ 
operator and thus with the same value for the one-particle $S$ matrix.
Thus, within the pseudofermion representation we again 
denote the above four states by $\vert c0,c0;-1\rangle$,
$\vert c0,c0;0\rangle$, $\vert  c0,c0,c1;0\rangle$, and 
$\vert c0,c0;+1\rangle$ where the branch indices refer to
the quantum objects created under the corresponding ground-state - 
excited-state transition. These four many-pseudofermion states
correspond to four one-pseudofermion scattering states whose
scatterer is a $c0$ pseudofermion hole. 
For electronic densities $n<1$
one has that the index with values $0,\pm 1$ refers to the 
$\eta$-spin projection deviation $\Delta S_{c}^{z}$ of the excited 
states, rather than to $S_{c}^{z}$. Indeed, the $n<1$ initial ground
state has a finite value for the $\eta$-spin projection. Another
important difference is that the $c1$ pseudofermion scattering
center of the excited state $\vert  c0,c0,c1;0\rangle$ can be created
under the ground-state - excited-state transition for
bare-momentum values in the range 
$q_3\in [-(\pi -2k_F),+(\pi -2k_F)]$ and has a $q_3$ dependent
energy dispersion, plotted in Figs. 8 and 9 of Ref. \cite{II}.

At fixed values of $q_1$, $q_2$, and $q_3$ there are for the extended 
spinon-holon theory four one-particle scattering states which
correspond to four well-defined excited many-particle states. As in Sec. III
we denote the latter four states by $\vert +1/2,+1/2;c\rangle$, 
$\vert +1/2,-1/2;c\rangle$, $\vert -1/2,+1/2;c\rangle$, 
and $\vert -1/2,-1/2;c\rangle$. They correspond to the above four excited 
energy eigenstates. (We recall that the 
relation of such bare-momentum values to the quantum numbers of the equations 
introduced by Takahashi \cite{Takahashi} is defined by Eqs. (A.1) and (B.1) of 
Ref. \cite{I}.)  The $n<1$ initial ground state has a finite occupancy of holons 
corresponding to $k$ values of the BA distribution of $k's$ excitation spectrum
in the range  $Q<\vert k\vert <\pi$. However, the two holons
associated with the $\eta$-spin state indices of these excited 
states are those created under the ground-state - excited-state transitions
at $k_1 =k^0 (q_1)$ and $k_2 =k^0 (q_2)$ in the range $k_1,\,k_2\in [-Q,\,+Q]$ 
with $k_1\neq k_2$.  Here $k^0 (q)$ is the rapidity function defined by the
first equation of Eq. (A.1) of Ref. \cite{V-1} and the $k$ {\it Fermi} value $Q$ 
is the parameter introduced in Ref. \cite{Lieb}, which 
is related to the $c0$ bare-momentum  {\it Fermi}
value $2k_F$ by Eq. (A.5) of Ref. \cite{V-1}. As for $n=1$, one finds that the four 
many-particle excited states associated with the holon-holon one-particle 
scattering states of the extended spinon-holon theory have in terms of the 
corresponding excited energy eigenstates associated with the 
one-pseudofermion scattering states expressions similar to those
of Eq. (\ref{scstates-U}) for $\alpha =c$, $\alpha\nu =c0$, and $\alpha\nu' =c1$.
For initial ground states with electronic density $n<1$ the excited 
energy eigenstates $\vert c0,c0;0\rangle$ and $\vert  c0,c0,c1;0\rangle$ 
have not the same energy and thus the excited states 
$\vert +1/2,-1/2;c\rangle$ and $\vert -1/2,+1/2;c\rangle$ 
associated with the holon-holon one-particle scattering states are not 
energy eigenstates. Moreover, for $n<1$ initial ground states
the energy and momentum expectation values of these
two excited states are different from the energy and momentum of the
first and fourth excited states of Eq. (\ref{scstates-U}). Indeed,
for $n<1$ there is no $\eta$-spin $SU(2)$ 
rotation symmetry, in contrast to the $n=1$ case considered
in Refs. \cite{S0,S}. It follows that for $n<1$ the energy and momentum 
expectation values of the second and third excited states of 
Eq. (\ref{scstates-U}) are not determined by the energy and 
momentum values of the two involved holons only: the length-one charge 
rapidity also contributes to these expectation values through its energy
dispersion, which is a function of the bare-momentum value $q_3$. 
(See Figs. 8 and 9 of Ref. \cite{II}.)
In contrast, for the pseudofermion scattering theory the $c1$ pseudofermion 
created under the ground-state - excited-state transition
is an independent scattering center and scatterer in its own right,
just as the two created $c0$ pseudofermion holes. Indeed, for
the pseudofermion representation the excited energy
eigenstate denoted here by $\vert  c0,c0,c1;0\rangle$ 
also contains a one-pseudofermion state whose
scatterer is the $c1$ pseudofermion.

A similar analysis could be performed for a generalization of the 
spin triplet and singlet excited states considered in Sec. III
(two-spinon states, within the extended spinon-holon representation) 
with a $m> 0$ initial ground state, 
as well as for any other PS excited states involving the creation
of a finite number of $\alpha\nu\neq c0,\,s1$ pseudofermions.

Finally, let us discuss the suitability for applications to the study
of the finite-energy spectral and dynamical properties of the two
alternative scattering theories.  For the $n<1$ metallic phase it is
desirable for the study of these properties that all one-particle 
scattering states correspond to energy eigenstates. This allows 
the use of suitable Lehmann representations for the spectral 
functions \cite{V-1,LE}. However, only for the reduced subspace 
considered in Sec. III corresponding to the $n=1$ Mott-Hubbard 
insulator initial ground state, all one-particle scattering 
states of the extended spinon-holon theory refer to energy eigenstates.  
Unfortunately, for initial ground states with electronic
density in the range $0<n<1$ (and spin density in the range
$0<m<n$) there are for such
an extended theory many one-particle scattering states which 
do not refer to energy and momentum eigenstates. Thus, Lehmann 
representations for the spectral functions as those used in the PDT 
of Refs. \cite{V-1,LE} cannot be used for the metallic phase in the 
case of the extended spinon-holon scattering theory.  In contrast, 
such a problem does not occur for the one-pseudofermion scattering 
states, which for the whole PS and all density values
always correspond to excited energy and momentum eigenstates. 
For the $n=1$ and $m=0$ initial ground state of the reduced 
subspace considered in Sec. III the one-particle scattering states 
of the spinon-holon representation refer to energy eigenstates. 
However, since the scatterers and scattering centers of that theory 
have $\eta$-spin $1/2$ or spin $1/2$, the $SO(4)$ symmetry implies 
that the $S$ matrix has a Yang Baxter Equation (YBE) like factorization, 
as the BA bare $S$ matrix of the original spin $1/2$ electrons, instead 
of the stronger commutative factorization of the pseudofermion 
and hole $S$ matrix. The $S$ matrix (\ref{MS}) has indeed such a 
property \cite{Natan,S0,S}.

Another advantage of the pseudofermion scattering theory of 
Refs. \cite{S-P0,S-P} for applications to the study of the dynamical 
properties is that the $S$ matrix of its scatterers has dimension one. 
Let us consider a $\alpha\nu$ pseudofermion scatterer of canonical 
momentum ${\bar{q}}$ and a $\alpha'\nu'$ pseudofermion scattering 
center of canonical momentum ${\bar{q}'}$. Thus, the 
canonical-momentum values ${\bar{q}}$ and ${\bar{q}'}$ correspond to 
an ``out'' state and a virtual state, respectively.  The 
corresponding pseudofermion anticommutators read \cite{S-P},
\begin{equation}
\{f^{\dag }_{{\bar{q}},\,\alpha\nu},\,f_{{\bar{q}}',\,\alpha'\nu'}\} =
{\delta_{\alpha\nu,\,\alpha'\nu'}\over N^*_{\alpha\nu}}\,\Bigl[S_{\alpha\nu}
(q)\Bigr]^{1/2}\,e^{-i({\bar{q}}-{\bar{q}}')/ 2}\,{{\Im}\Bigl(\Bigl[S_{\alpha\nu}
(q)\Bigr]^{1/2}\Bigr)\over\sin ([{\bar{q}}-{\bar{q}}']/2)} \, ; \hspace{0.25cm} \{f^{\dag
}_{{\bar{q}},\,\alpha\nu},\,f^{\dag
}_{{\bar{q}}',\,\alpha'\nu'}\}=\{f_{{\bar{q}},\,\alpha\nu},\,f_{{\bar{q}}',\,\alpha'\nu'}\}=0\, . 
\label{pfacrGS-S}
\end{equation}
Note that the first pseudofermion anticommutation relation can be
expressed solely in terms of the difference $[{\bar{q}}-{\bar{q}}']$ and 
the $S$ matrix of the excited-state $\alpha\nu$ pseudofermion scatterer. 
Following the results of Ref. \cite{V-1}, the one- and two-electron matrix 
elements between the initial ground state and the excited energy 
eigenstates can be expressed in terms of the anticommutators 
(\ref{pfacrGS-S}). Thus, within the pseudofermion representation the 
$S$ matrix $S_{\alpha\nu} (q_j)$ given in Eq. (\ref{San}) controls the 
spectral properties of the model. If it had dimension larger than one, 
the problem would be much more involved. This is the case of the 
spinon-holon scattering theory of Refs. \cite{Natan,S0,S}, whose 
scatterers and scattering centers are spin $1/2$ spinons and 
$\eta$-spin $1/2$ holons. As shown in Sec. III for the reduced subspace,
the corresponding spinon-spinon and holon-holon $S$ matrices are 
indeed non-diagonal and thus the problem of the evaluation
of these matrix elements is much more complex for the spinon-holon
representation. 

Such a problem simplifies for the pseudofermion representation because 
the PS subspaces associated with a given one- or two-electron
spectral function can be expressed in terms of direct products 
corresponding to each of the $\alpha\nu$ pseudofermion occupancy 
configurations of branches with finite pseudofermion occupancy \cite{V-1,LE}. 
For these matrix elements the direct product is associated 
with the commutative factorization of the $S$ matrices
provided in Eq. (\ref{San}) in terms of the elementary $S$ matrices 
$S_{\alpha\nu ,\,\alpha'\nu'} (q_j, q_{j'})$, Eq. (\ref{Sanan}). 
Such commutativity is stronger than the symmetry 
associated with the YBE and results from the elementary $S$ matrices
$S_{\alpha\nu ,\,\alpha'\nu'} (q_j, q_{j'})$ being simple phase factors, instead of
matrices of dimension larger than one. The commutative factorization of the 
$S$ matrix occurs when the one-particle scattering states correspond
to energy eigenstates and the scatterers and scattering centers are 
$\eta$-spin-neutral and/or spin-neutral, as occurs for the pseudofermion scattering
theory \cite{S-P}.  Unfortunately, for initial ground states with densities in 
the ranges $0<n<1$ and $0<m<n$ many one-particle scattering states 
of the extended spinon-holon theory do not refer to energy eigenstates. For
these states the occupancy configurations of the BA charge (and spin) string 
excitations of length $\nu =1,2,...$ (and $\nu =2,3,...$) are
included in the holon (and spinon) scatterers and scattering centers so 
that the corresponding BA string branches lose their independent character.
Thus, for the extended spinon-holon theory the above PS subspaces are 
expressed as the direct product of two subspaces only, referring 
to the holon and spinon occupancy configurations, respectively. 
It is this property of the extended spinon-holon 
theory that increases the complexity 
of the evaluation of the spectral functions for the metallic phase. Indeed, 
such a direct product does not include the BA charge and string excitations
of length $\nu$ as independent branches, corresponding to 
independent scatterers and scattering centers. In contrast, within the pseudofermion
representation such BA charge (and spin) string excitations
of length $\nu =1,2,...$ (and $\nu =2,3,...$) refer to independent 
$c\nu$ pseudofermion-scatterer (and $s\nu$ pseudofermion-scatterer) branches
which exist in their own right. 
The corresponding $c\nu$ pseudofermion (and $s\nu$ pseudofermion) 
occupancy configurations refer to independent subspaces 
which contribute to the direct product of 
the whole PS subspace relevant for the spectral function
under consideration. 

Last but not least, the holons (and spinons) of the
extended spinon-holon representation
always involve the quantum superposition of the degrees of freedom
associated with the $c0$ pseudofermions and Yang holons
(and $s1$ pseudofermions and HL spinons), which are not
invariant and are invariant under the electron - rotated-electron
unitary transformation, respectively. (When the holon-holon
or spinon-spinon one-particle scattering states of the extended 
spinon-holon theory do not refer to energy eigenstates,
the corresponding holons or spinons also involve the $c\nu\neq c0$ 
pseudofermion or $s\nu\neq s1$ pseudofermion occupancy
configurations, respectively, as discussed above.)
It follows that the holons (and spinons) of the extended spinon-holon 
representation are not invariant under that transformation.
It turns out that the electron - rotated-electron unitary transformation
plays a major role in the PDT of Refs. \cite{V-1,LE}. For
instance, the contribution of the Yang holons and HL spinons to the 
evaluation of the spectral functions by the PDT is considerably
simplified by their invariance under that 
transformation. In contrast, the holon and spinon definition
of the extended spinon-holon representation 
does not profit from the symmetries
associated with such a unitary transformation, which renders impossible
the use of key PDT procedures for the evaluation of the
finite-energy one-electron and two-electron spectral functions. 

%%%%%%%%%%%%%%%%%%%%%%%%%%%%%%%%%%%%%%%%%%%%%%%%%%%%%%%%%%%%%%%%
\section{CONCLUDING REMARKS}

The quantum objects associated with the pseudofermion representation of Refs. \cite{I,S-P0,S-P} 
emerge naturally from the electron - rotated-electron unitary transformation. 
Such a transformation was shown in Ref. \cite{I} to correspond to the first
step performed by the exact diagonalization of the non-perturbative many-electron
quantum problem. Therefore, the choice of quantum objects of 
Refs. \cite{I,S-P0,S-P} profits from the symmetries
associated with the electron - rotated-electron unitary transformation. For instance,
the holons and spinons are defined in such away that they either remain invariant
under that transformation (Yang holons and HL spinons) but then do not scatter
or do not remain invariant under the same transformation and thus cannot
exist as independent quantum objects. 

Indeed, the holons (and spinons) that are not invariant under the electron - rotated-electron 
unitary transformation are always part of $2\nu$-holon (and $2\nu$-spinon) composite
$\eta$-spin singlet (and spin singlet) pseudofermions, where $\nu =1,2,...$ gives the number 
of pairs of $+1/2$ holons and $-1/2$ holons (and $+1/2$ spinons and $-1/2$ spinons). 
Interestingly, in the pseudofermion scattering theory the relation of the composite
$c\nu$ pseudofermion (and $s\nu$ pseudofermion) scatterers and scattering centers to the holons 
(and spinons) has similarities with that of the physical 
particles to the quarks in chromodynamics \cite{Martinus}. Within the latter theory all quark
composite physical particles must be color-neutral, yet the quarks have color. On the other
hand, in the pseudofermion scattering theory all $2\nu$-holon (and $2\nu$-spinon) 
composite pseudofermion scatterers and scattering centers must have zero $\eta$-spin
(and spin) and thus must be $\eta$-spin-neutral 
(and spin-neutral), yet the holons (and spinons) have finite $\eta$-spin $1/2$ 
(and spin $1/2$). (The $c0$ pseudofermion scatterers and scattering centers are
not composed of holons or spinons but are $\eta$-spin-less and spin-less objects.) 
In turn, the Yang holons and HL spinons have finite $\eta$-spin $1/2$ 
and spin $1/2$, respectively, but do not scatter. 

As discussed above, it is precisely the $\eta$-spin-neutral 
(and spin-neutral)  character of the $2\nu$-holon (and $2\nu$-spinon) composite pseudofermion 
scatterers and scattering centers  and the $\eta$-spin-less and spin-less character
of the $c0$ pseudofermion scatterers and scattering centers which is behind the
dimension of their $S$ matrix $S_{\alpha\nu} (q_j)$ given in Eq. (\ref{San}). 
We emphasize that such a $S$ matrix fully controls 
the pseudofermion anticommutators through Eq. (\ref{pfacrGS-S}) and also the 
value of the matrix elements between energy eigenstates and the corresponding
finite-energy spectral properties, as confirmed by the studies of Refs. \cite{V-1,LE}.
Furthermore, within the pseudofermion representation all one-pseudofermion
scattering states correspond to energy and momentum eigenstates.

Our study of the relation between the many-particle states associated
with the one-particle scattering states of the conventional spinon-holon
representation \cite{Natan,S0,S} and pseudofermion description 
\cite{I,S-P0,S-P}, respectively, 
reveals that the $\eta$-spin $1/2$ holon and spin $1/2$ spinon scatterers and 
scattering centers of the former theory are different from the pseudofermion 
and pseudofermion-hole scatterers and scattering centers. The construction of 
the holon and spinon scatterers and scattering centers
of the spinon-holon representation does not profit from the symmetries
associated with the electron - rotated-electron unitary transformation. For
instance, the holon and spinon scatterers and scattering centers 
of that theory involve mixing of the quantum 
objects of the pseudofermion representation that are not
invariant under that transformation. It follows that many of the
one-particle scattering states of the spinon-holon theory 
do not refer to $\eta$-spin and
spin eigenstates and thus the corresponding holon-holon and spinon-spinon
$S$ matrices are not diagonal. Moreover, for metallic initial
ground states many of the one-particle scattering states of the extended
spinon-holon theory do not correspond to energy and 
momentum eigenstates. As discussed in the previous section, 
these features of the extended spinon-holon representation  
imply that its use in the study of the finite-energy 
spectral and dynamical properties of the metallic phase is 
a much more involved problem than the use of the
pseudofermion scattering theory to study such properties.

Our investigation also reveals that both representations
are faithful and thus that
there is no inconsistency between the two corresponding definitions of quantum
objects. The problem clarified in this paper is of interest for the 
further understanding of the unusual spectral properties observed
in low-dimensional complex materials. Indeed, by use of the PDT of Refs. \cite{V-1,LE}, the 
unusual independent charge and spin finite-energy spectral features observed recently 
by angle-resolved photoelectron spectroscopy in quasi-1D metals \cite{spectral0} 
were shown in Ref. \cite{spectral}  to correspond to charge $c0$ and spin $s1$,
respectively, pseudofermion-hole scatterers and scattering centers of the type 
considered in Refs. \cite{S-P0,S-P}. (That such features correspond to charge
and spin pseudofermions can be proved from the form of the $S$ matrix used in 
the evaluation of the one-electron matrix elements between the ground state and 
the excited states.)  Thus, the exotic scatterers and scattering centers studied here 
and in these references exist in real low-dimensional complex 
materials. This indeed justifies the interest of clarifying their relation to the quantum
objects of the conventional spinon-holon representation of Refs. \cite{Natan,S0,S,TS}, 
which have been used in many theoretical studies of low-dimensional 
electronic correlated problems.

%%%%%%%%%%%%%%%%%%%%%%%%%%%%%%%%%%%%%%%%%%%%%%%%%%%%%%%%%%%%%%%%%%%%%%%%%%
\begin{acknowledgments}
We thank Daniel Bozi, Vladimir E. Korepin, Patrick A. Lee, Sung-Sik Lee, Karlo Penc, and Pedro D.
Sacramento for stimulating discussions. JMPC thanks the hospitality
of MIT, where part of this research was performed, and the support of the ESF Science
Programme INSTANS 2005-2010 and FCT grant POCTI/FIS/58133/2004. KEH thanks the support of
FCT under the grant SFRH/BPD/11513/2002.
\end{acknowledgments}
%%%%%%%%%%%%%%%%%%%%%%%%%%%%%%%%%%%%%%%%%%%%%%%%%%%%%%%%%%%%%%%%%%%%%%%%%%
\appendix

\section{USEFUL PHASE-SHIFT EXPRESSIONS}

Here we provide expressions for phase shifts and other quantities 
needed for the evaluation of the overall phase shift given in 
Eq. (\ref{Qcan1j}). We start by providing the general expression for 
the overall scatter-less phase shift $Q_{\alpha\nu}^0$ on the 
right-hand side of Eq. (\ref{Qcan1j}). It is given by \cite{S-P}, 
\begin{eqnarray}
Q_{c0}^0 & = & 0 \, ; \hspace{0.5cm} \sum_{\alpha =c,\,s}\,\sum_{\nu=1}^{\infty} \Delta
N_{\alpha\nu} \hspace{0.25cm} {\rm even} \, ;  \hspace{1.0cm} Q_{c0}^0=\pm\pi \, ;
\hspace{0.5cm} \sum_{\alpha =c,\,s}\,\sum_{\nu=1}^{\infty} \Delta
N_{\alpha\nu} \hspace{0.25cm} {\rm odd} \, ; \nonumber \\
Q_{\alpha\nu}^0 & = & 0 \, ; \hspace{0.5cm} \Delta N_{c0}+\Delta N_{\alpha\nu}
\hspace{0.25cm} {\rm even} \, ; \hspace{1.0cm} Q_{\alpha\nu}^0=\pm\pi \, ; \hspace{0.5cm}
\Delta N_{c0}+\Delta N_{\alpha\nu} \hspace{0.25cm} {\rm odd} \, ; \hspace{0.5cm} \alpha =
c,\,s \, , \hspace{0.25cm} \nu > 0 \, . \label{pic0an}
\end{eqnarray}
When $Q_{\alpha\nu}^0=\pm\pi$ for the $\alpha\nu\neq c0$ bands, the uniquely
chosen and only permitted value $Q_{\alpha\nu}^0=\pi$ or $Q_{\alpha\nu}^0=-\pi$ is that which
leads to symmetrical limiting discrete bare-momentum values $\pm [\pi/L][N^*_{\alpha\nu}-1]$
for the excited-state bare-momentum band. (See Eq. (B.14) of Ref. \cite{I}.)
In turn, for the $c0$ branch the bare-momentum band width is $2\pi$. Thus, in
this case $Q_{c0}^0=\pi$ and $Q_{c0}^0=-\pi$ lead to allowed occupancy configurations
of alternative excited energy eigenstates. (In the particular case that the $c0$
band is full for the excited energy eigenstate, the two values $Q_{c0}^0=\pi$ and 
$Q_{c0}^0=-\pi$ refer to two equivalent representations of that state.)

In the remaining of this Appendix we provide quantities needed for the derivation
of the overall phase-shift expressions (\ref{Qc-eT})-(\ref{Qc-s1-1-eD}).
We start by providing the bare-momentum distribution function deviations for
all types of excited states considered in Tables I and II of Sec. III. Use
of such deviations in the general expression for the overall phase shift given in 
Eq. (\ref{Qcan1j}) leads straightforwardly to the phase-shift expressions
(\ref{Qc-eT})-(\ref{Qc-s1-eD}). The three classes of $\eta$-spin triplet 
excited energy eigenstates considered in that section have the same 
expression for such deviations given by,
\begin{equation}
\Delta N_{c0} (q) = -{2\pi\over L}\sum_{l=1}^{2}\delta (q-q_l) -{\pi\over L}\,\delta
(q+\pi)+{\pi\over L}\,\delta (q-\pi) \, , \hspace{0.25cm} \vert q_l\vert\leq\pi 
\, ; \hspace{0.50cm}
\Delta N_{s1} (q) = -{\pi\over L}\,\delta (q+\pi/2)-{\pi\over L}\,\delta (q-\pi/2) \, .
\label{DNc-eT}
\end{equation}
The deviation $\Delta N_{s1} (q)$ given here also applies to the $\eta$-spin singlet excited 
states, whereas for the $c0$ and $c1$ branches the deviations read as follows for these 
states,
\begin{equation}
\Delta N_{c0} (q) = -{2\pi\over L}\sum_{l=1}^{2}\delta (q-q_l)  \, ; \hspace{0.50cm} \Delta N_{c1} (q) =
\delta_{q,\,0} \, . \label{DNc0s1c1-eS}
\end{equation}
The three classes of spin triplet excitations of Sec. III have again the same bare-momentum 
distribution function deviations given by,
\begin{equation}
\Delta N_{c0} (q) = -{\pi\over L}\,\delta (q+\pi)+{\pi\over L}\,\delta (q-\pi) \, ; \hspace{0.50cm}
\Delta N_{s1} (q) = -{2\pi\over L}\sum_{l=1}^{2}\delta (q-{q'}_l) +{\pi\over L}\,\delta
(q+\pi/2)+{\pi\over L}\,\delta (q-\pi/2) \, , \hspace{0.25cm} \vert {q'}_l\vert\leq\pi/2
\, . \label{DNc-T}
\end{equation}
The deviation $\Delta N_{c0} (q)$ given here also applies to the spin singlet excited 
states, whereas for the $s1$ and $s2$ branches the deviations read as follows for these 
states,
\begin{equation}
\Delta N_{s1} (q) = -{2\pi\over L}\sum_{l=1}^{2}\delta (q-{q'}_l) \, ,
\hspace{0.25cm} \vert q_l\vert\leq\pi \, ; \hspace{0.50cm} \Delta N_{s2} (q) =
\delta_{q,\,0} \, . \label{DNc0s1s2-S}
\end{equation}
The four classes of doublet excited states have the same bare-momentum
distribution function deviations given by,
\begin{equation}
\Delta N_{c0} (q) = -{2\pi\over L}\delta (q-q_1) \, , \hspace{0.25cm} \vert
q_1\vert\leq\pi \, ; \hspace{0.50cm} \Delta N_{s1} (q) = -{2\pi\over L}\delta (q-{q'}_1)
\, , \hspace{0.25cm} \vert {q'}_1\vert\leq \pi/2 \, . \label{DNcs1-eD}
\end{equation}

Finally, we provide several two-pseudofermion phase-shift expressions needed
for the derivation of the overall phase-shift expressions given in Eqs.
(\ref{Qc-1-eT})-(\ref{Qc-s1-1-eD}). The rapidity two-pseudofermion 
phase shifts $2\pi\,{\bar{\Phi }}_{\alpha\nu,\,\alpha'\nu'}(r,r')$ are defined 
by the integral equations (A1)-(A13) of Ref. \cite{IIIb}. We solve these 
equations by Fourier transforming them after considering that $Q=\pi$ 
and $B=\infty$ and thus $r^0_c = 4t\,\sin Q/U=0$ and $r^0_s = 4t\,B/U=\infty$ 
for finite values of $U/t$. Such a procedure leads to the following 
expressions valid for $n\rightarrow 1$, $m\rightarrow 0$, and
finite values of $U/t$ for the two-pseudofermion phase shifts involving the $c0$ 
and $s1$ scatterers,
\begin{equation}
2\pi\,{\bar{\Phi }}_{c0,\,c0}(r,\,r') = - 2B (r-r') \, ; \hspace{0.5cm} 2\pi\,{\bar{\Phi
}}_{c0,\,s1}(r,\,r') = -{\rm arc}\tan\Bigl(\sinh \Bigl({\pi\over
2}(r-r')\Bigr)\Bigr) \, , \label{Phicc-cs1}
\end{equation}
\begin{eqnarray}
2\pi\,{\bar{\Phi }}_{s1,\,c0}(r,\,r') & = & -{\rm arc}\tan\Bigl(\sinh
\Bigl({\pi\over 2}(r-r')\Bigr)\Bigr) \, ;
\hspace{0.5cm} r \neq \pm \infty  \nonumber \\
& = & -{{\rm sgn} (r)\pi\over \sqrt{2}} \, ; \hspace{0.5cm} r = \pm \infty \, ,
\label{Phisc1}
\end{eqnarray}
\begin{eqnarray}
2\pi\,{\bar{\Phi }}_{s1,\,s1}(r,\,r') & = & 2B (r-r') \, ; \hspace{0.3cm} r\neq\pm\infty  
\nonumber
\\
& = & {{\rm sgn} (r)\pi\over \sqrt{2}} \, ; \hspace{0.5cm} r = \pm\infty \, ,
\hspace{0.5cm} r' \neq r \nonumber
\\
& = & [{\rm sgn} (r)]\Bigl({3\over \sqrt{2}}-2\Bigr)\pi \, ; \hspace{0.5cm} r = r' =
\pm\infty \, , \label{Phiss1}
\end{eqnarray}
\begin{equation}
2\pi\,{\bar{\Phi }}_{c0,\,c1}(r,\,r') = -2\,{\rm arc}\tan (r-r') \, , \label{Phicn-csn1}
\end{equation}
\begin{eqnarray}
2\pi\,{\bar{\Phi }}_{s1,\,s2}(r,\,r') & = & 2\,{\rm arc}\tan (r-r') \, ; \hspace{0.5cm} r\neq \pm\infty \, , 
\nonumber \\
& = & \pm {2\pi\over\sqrt{2}} \, ; \hspace{0.5cm} r= \pm\infty \, , \label{Phicn-ssn1}
\end{eqnarray}
and $2\pi\,{\bar{\Phi }}_{c0,\,s2}(r,\,r') =2\pi\,{\bar{\Phi }}_{s1,\,c1}(r,\,r') =  0$. Here,
\begin{equation}
2B (r) = i \ln {\Gamma \Bigl({1\over 2}+i{r\over 4}\Bigr)\,\Gamma
\Bigl(1-i{r\over 4}\Bigr)\over \Gamma \Bigl({1\over 2}-i{r\over 4}\Bigr)\,\Gamma
\Bigl(1+i{r\over 4}\Bigr)} \, . \label{Br}
\end{equation}
For the two types of excited states of Table I of Sec. III with one 
$c1$ pseudofermion and one $s2$ pseudofermion, respectively,
the $n\rightarrow 1$ and 
$m\rightarrow 0$ equations $Q^{\Phi}_{c1} (0)=0$ and $Q^{\Phi}_{s2} (0)=0$
associated with the scatter-less character of such a pseudofermion
can be written as 
$\sum_{l=1}^{2}\,{\rm arc}{\rm tan}(4t[\Lambda_{c1}(0)-
\Lambda^{0}_{c0}(q_l)]/U) =
0$ and  $\sum_{l=1}^{2}\,{\rm arc}{\rm tan}(4t[\Lambda_{s2}(0)- 
\Lambda^{0}_{s1}({q'}_l)]/U) = 0$, respectively \cite{S-P}.
Solution of these equations leads to 
$\Lambda_{c1}(0)=\Lambda_{c1}(0,q_1,q_2)=
[\Lambda^{0}_{c0}(q_1)+\Lambda^{0}_{c0}(q_2)]/2$ and
$\Lambda_{s2}(0)=\Lambda_{s2}(0,{q'}_1,{q'}_2)=
[\Lambda^{0}_{s1}({q'}_1)+\Lambda^{0}_{s1}({q'}_2)]/2$,
respectively. Use of that solution in the rapidity two-pseudofermion expressions 
of Eqs. (\ref{Phicn-csn1}) and (\ref{Phicn-ssn1}) leads then to the 
following expressions for the two-pseudofermion phase shifts 
$2\pi\,\Phi_{c0,\,c1}(q_1,0)$ and $2\pi\,\Phi_{s1,\,s2}({q'}_1,0)$
for ${q'}_1\neq\pm\pi/2$,
\begin{eqnarray}
2\pi\,\Phi_{c0,\,c1}(q_1,0) & = & 2\pi\,\bar{\Phi }_{c0,\,c1}
\Bigl({4t\,\Lambda^0_{c0}(q_1)\over U}, {4t\,\Lambda_{c1}(0,q_1,q_2)\over U}\Bigr) =
-2\,{\rm arc}{\rm tan}\Bigl({2t\,[\Lambda^0_{c0}(q_1)-\Lambda^0_{c0}(q_2)]\over
{U}}\Bigr)\, ; \nonumber \\
2\pi\,\Phi_{s1,\,s2}({q'}_1,0) & = & 2\pi\,\bar{\Phi }_{s1,\,s2}
\Bigl({4t\,\Lambda^0_{s1}({q'}_1)\over U}, {4t\,\Lambda_{s2}(0,{q'}_1,{q'}_2)\over U}\Bigr) =
2\,{\rm arc}{\rm tan}\Bigl({2t\,[\Lambda^0_{s1}({q'}_1)-\Lambda^0_{s1}({q'}_2)]\over
{U}}\Bigr) \, . \label{Phis-2}
\end{eqnarray}

%%%%%%%%%%%%%%%%%%%%%%%%%%%%%%%%%%%%%%%%%%%%%%%%%%%%%%%%%%%%%%%%%%%%%%%%%%


\begin{references}
\bibitem[1]{spectral0}
        R. Claessen, M. Sing,
        U. Schwingenschl\"ogl, P. Blaha, M. Dressel, and C. S. Jacobsen,
        Phys. Rev. Lett. {\bf 88} 096402 (2002); M. Sing, U. Schwingenschl\"ogl, 
        R. Claessen, P. Blaha, J. M. P. Carmelo, L. M. Martelo, P. D. Sacramento, M.
        Dressel, and C. S. Jacobsen, Phys. Rev. B {\bf 68}, 125111 (2003).
\bibitem[2]{spectral}
        J. M. P. Carmelo, K. Penc, L. M. Martelo, P. D. Sacramento,
        J. M. B. Lopes dos Santos, R. Claessen, M. Sing, and
        U. Schwingenschl\"ogl, Europhys. Lett. {\bf 67}, 233 (2004);
        J. M. P. Carmelo, D. Bozi, P. D. Sacramento, and K. Penc,
        submitted for publication.
\bibitem[3]{Eric}
        H. Benthien, F. Gebhard, and E. Jeckelmann, Phys. Rev.
        Lett. {\bf 92}, 256401 (2004).
\bibitem[4]{Lieb}
        Elliott H. Lieb and F. Y. Wu, Phys. Rev. Lett. {\bf 20}, 1445 (1968).
\bibitem[5]{Takahashi}
        M. Takahashi, Prog. Theor. Phys. {\bf 47}, 69 (1972);
        J. M. P. Carmelo and N. M. R. Peres, Phys. Rev. B {\bf 56}, 3717 (1997).
\bibitem[6]{Zoller}
        D. Jaksch and P. Zoller, Ann. Phys. {\bf 315}, 52 (2005).
\bibitem[7]{Chen}
        Y. Chen, private communication.
\bibitem[8]{Woy}
        F. Woynarovich, J. Phys. A: Math. Gen. {\bf 22}, 4243 (1989).
\bibitem[9]{Ogata}
        M. Ogata and H. Shiba, Phys. Rev. B {\bf 41}, 2326 (1990);
        M. Ogata, T. Sugiyama, and H. Shiba, Phys. Rev. B {\bf 43}, 8401 (1991).
\bibitem[10]{Kawakami}
        N. Kawakami and S. K. Yang, Phys. Lett. A {\bf 148}, 359 (1990).
\bibitem[11]{Frahm}
        H. Frahm and V. E. Korepin, Phys. Rev. B {\bf 42}, 10 553
        (1990); H. Frahm and V. E. Korepin, Phys. Rev. B {\bf 43}, 5653 (1991).
\bibitem[12]{Brech}
        M. Brech, J. Voit, and H. Buttner, Europhys. Lett. {\bf 12}, 289 (1990).
\bibitem[13]{Karlo}
        K. Penc and J. S\'olyom, Phys. Rev. B {\bf 47}, 6273 (1993).
\bibitem[14]{93-94}
        J. M. P. Carmelo and A. H. Castro Neto, Phys. Rev. Lett. {\bf 70}, 1904 (1993); 
        J. M. P. Carmelo, A. H. Castro Neto, and D. K. Campbell, Phys. Rev. B {\bf 50}, 
        3667 (1994) 3667; J. M. P. Carmelo, A. H. Castro Neto, and D. K. Campbell, 
        Phys. Rev. B {\bf 50}, 3683 (1994).
\bibitem[15]{V-1}
        J. M. P. Carmelo, K. Penc, and D. Bozi, Nucl. Phys. B {\bf 725},
        421 (2005); J. M. P. Carmelo, K. Penc, and D. Bozi, Nucl. Phys. B {\bf 737}, 
        351 (2006), Erratum; J. M. P. Carmelo and K. Penc, cond-mat/0311075.
\bibitem[16]{LE}
        J. M. P. Carmelo, L. M. Martelo, and K. Penc, Nucl. Phys. B  {\bf 737},
        237 (2006); J. M. P. Carmelo and K. Penc, at press in Phys. Rev. B (2006)
\bibitem[17]{BS}
        H. Basista, D. A. Bonn, T. Timusk, J. Voit, D. J\'erome,
        K. Bechgaard, Phys. Rev. B {\bf 42}, 4088 (1990);
        J. M. P. Carmelo, P. Horsch, D. K. Campbell, and A. H. Castro Neto,
        Phys. Rev. B (RC) {\bf 48}, 4200 (1993).
\bibitem[18]{S-P0}
        J. M. P. Carmelo, J. Phys.: Cond. Mat. {\bf 17}, 5517 (2005).
\bibitem[19]{S-P}
        J. M. P. Carmelo, D. Bozi, and P. D. Sacramento, cond-mat/06....
\bibitem[20]{properties}
        Dionys Baeriswyl, Jos\'e Carmelo, and Kazumi Maki, 
        Synth. Met. {\bf 21}, 271 (1987); N. M. R. Peres, J. M. P. Carmelo,
        D. K. Campbell, and A. W. Sandvik, Z. Phys. B {\bf 103}, 217 (1997);
        J. M. P. Carmelo, N. M. R. Peres, and P. D.
        Sacramento, Phys. Rev. Lett. {\bf 84}, 4673 (2000);
        D. Controzzi, F.H.L. Essler, and A.M. Tsvelik,
        Phys. Rev. Lett. {\bf 86}, 680 (2001); 
        H. Kishida, M. Ono, K. Miura, H. Okamoto, M. Izumi, T. Manako,
        M. Kawasaki, Y. Taguchi, Y. Tokura, T. Tohyama, K. Tsutsui, and
        S. Maekawa, Phys. Rev. Lett. {\bf 87}, 177401 (2001).
\bibitem[21]{super}
        J. M. P. Carmelo, F. Guinea, K. Penc, and P. D. Sacramento,
        Europhys. Lett. {\bf 68}, 839 (2004).
\bibitem[22]{Natan} N. Andrei, {\it Series on Modern Condensed Matter
        Physics - Vol. 6}, 458, World Scientific, Lecture Notes of
        ICTP Summer Course 1992, Editors: S. Lundquist, G. Morandi,
        and Yu Lu [cond-mat/9408101].
\bibitem[23]{S0}
        Fabian H. L. Essler and Vladimir E. Korepin,
        Phys. Rev. Lett. {\bf 72}, 908 (1994).
\bibitem[24]{S}
        Fabian H. L. Essler and Vladimir E. Korepin,
        Nucl. Phys. B {\bf 426}, 505 (1994).
\bibitem[25]{Natan0}
        N. Andrei and J. Lowenstein, Phys. Rev. Lett. {\bf 43}, 1698 (1979).
\bibitem[26]{NA80}  
	N. Andrei,  Phys. Rev. Lett. {\bf 45}, 379 (1980).
\bibitem[27]{Faddeev}
        L. D. Faddeev and L. A. Takhtajan, Phys. Lett. {\bf 85A}, 375
        (1981).
\bibitem[28]{HL}
        O. J. Heilmann and E. H. Lieb, Ann. N. Y. Acad. Sci. {\bf 172}, 583
        (1971); C. N. Yang, Phys. Rev. Lett. {\bf 63}, 2144 (1989).
\bibitem[29]{I}
        J. M. P. Carmelo, J. M. Rom\'an, and K. Penc, Nucl. Phys. B
        {\bf 683}, 387 (2004).
\bibitem[30]{IIIb}
        J. M. P. Carmelo, cond-mat/0405411.
\bibitem[31]{TS}
        P. W. Anderson, in
        {\it The Theory of Superconductivity in the High-$T_c$
        Cuprates} (PU Press, Princeton, NJ, 1997).
\bibitem[32]{II}
        J. M. P. Carmelo and P. D. Sacramento, Phys. Rev. B {\bf 68}, 085104 (2003).
\bibitem[33]{Penc}
        K. Penc, K. Hallberg, F. Mila, and H. Shiba,
        Phys. Rev. Lett. {\bf 77}, 1390 (1996);  K. Penc, K. Hallberg, F. Mila, and H. Shiba,
        Phys. Rev. B {\bf 55}, 15 475 (1997).
\bibitem[34]{PWA}
        J. C. Talstra and S. P. Strong, Phys. Rev. B {\bf 56}, 6094
        (1997).
\bibitem[35]{Tohyama}
        Y. Mizuno, K. Tsutsui, T. Tohyama, and S. Maekawa, Phys. Rev.
        B {\bf 62}, R4769 (2000).
\bibitem[36]{CFT}
        A. A. Belavin, A. M. Polyakov, and A. B.
        Zamolodchikov, Nucl. Phys. B {\bf 241}, 333 (1984).
\bibitem[37]{Bozo}
        H. J. Schulz, Phys. Rev. Lett. {\bf 64}, 2831 (1990);
        J. M. P. Carmelo, A. H. Castro Neto, and D. K. Campbell,
        Phys. Rev. Lett. {\bf 73}, 926 (1994) and Erratum {\bf 74}, 3089 (1995).
\bibitem[38]{tj}
        P. A. Bares, J. M. P. Carmelo, J. Ferrer, and
        P. Horsch, Phys. Rev. B {\bf 46}, 14 624 (1992).
\bibitem[39]{Korepin79}  V. E. Korepin, Teor. Mat. Fiz. {\bf 41}, 169 (1979);
        Theor. Math. Phys. {\bf 76}, 165 (1980); N. Andrei and J. H . Lowenstein, Phys. Lett. 
	{\bf 91B}, 401 (1980).
\bibitem[40]{Martinus} For a basic introduction to chromodynamics see Martinus Veltman, 
	{\em Facts and Mysteries in Elementary Particle Physics} (World Scientific, New
	Jersey, 2003).
\end{references}
\end{document}